\documentclass[journal=cgdefu,manuscript=article,layout=twocolumn]{achemso}

\usepackage[version=3]{mhchem} % Formula subscripts using \ce{}

\newcommand*\mycommand[1]{\texttt{\emph{#1}}}
\usepackage{xcolor}
\newcommand{\red}[1]{{\color{black}{#1}}}

\author{Lei Li}
\author{Jesus Rodriguez Sanchez}
\author{Felix Kohler}
\author{Anja R{\o}yne}
\author{Dag Kristian Dysthe}
\email{d.k.dysthe@fys.uio.no}
\phone{+47 90940996}
\affiliation[University of Oslo]
{Physics of Geological Processes (PGP), The NJORD Centre, Department of Physics, University of Oslo, PObox 1048 Blindern, 0316 Oslo}

\title{Microfluidic control of nucleation and growth of CaCO$_3$}

\abbreviations{IR,NMR,UV}

\begin{document}

\begin{abstract}
A novel method for studying nucleation and growth of CaCO$_3$ crystals
in situ has been developed and tested rigorously. We demonstrate that
precise flow control is essential and how this is achieved. The method
has the advantage that one may study single crystals of polymorphs
that are thermodynamically unstable in collections of many crystals
and that one obtains precise and accurate growth rates without any
extra assumptions. We also demonstrate that at low supersaturations
where 2D nucleation does not occur we measure the growth rate constant
of calcite to be \red{5 times} larger than that reported by batch methods
and \red{two orders of magnitude} larger than measured by AFM.  Considering the
large interest in calcite growth in for example geoscience,
environmental science, and industry we consider that it is important
to explain the discrepancy of growth rate constants between different
methods. \red{The method presented here can easily be applied to many other minerals.}
\end{abstract}

%\begin{keyword}
%:crystal growth rate; calcite; microfluidic; Elveflow stability; PDMS channel.
%\end{keyword}

\pagestyle{myheadings}
%\markboth{}{In preparation for Crystal Growth and Design}

\section{Introduction}
Calcium carbonate (CaCO$_3$) makes up about 4\% of the Earth's crust, it is the major long term sink in  the global carbon dioxide cycle~\cite{Banner1995,Riding2000,Fairchild2009}, 60\% of the worlds oil is held in carbonate reservoirs~\cite{Roehl1985} and it is probably the most studied biomineral~\cite{Watabe1981,Colfen2003,Meldrum2003}. Industrially it is used in paper, plastics, paints, coatings, personal health, food production, building materials and construction~\cite{Carr2000}. The modelling of biomineralization, geological, industrial and climate processes involving calcium carbonate relies on firm knowledge of CaCO$_3$ crystal growth mechanisms and rates~\cite{Bracco2013,Andersson2016}. Much effort has therefore been invested in such fundamental studies.~\cite{Reddy1981,Hillner1992,Stipp1994,Teng1998,Teng1999,Teng2000,Ruiz-Agudo2012}. Although
nanoscale measurements have revealed the fundamental mechanisms of
growth, the rates measured by atomic force microscopy (AFM)~\cite{Teng2000,Bracco2013} do not agree well with macroscopic methods~\cite{Reddy1981}. The purpose of this study is to develop a new technique at an
intermediate scale to study the nucleation and growth of whole
crystals. This allows direct measurement of surface normal growth rate
and dispels the need for additional measurements or assumptions.

Since the first \textit{in situ} AFM studies of
calcite dissolution and growth~\cite{Hillner1992} the AFM has been
invaluable in building our understanding of the calcite
surface~\cite{Ruiz-Agudo2012}.  \textit{In situ} AFM observations
allow the determination of step growth velocities of small portions of
a surface (preferably a well-defined screw
dislocation~\cite{Teng2000}). AFM studies of calcite
growth~\cite{Hillner1992,Stipp1994,Teng1998,Teng1999,Teng2000,Ruiz-Agudo2012}
give detailed input for microkinetic
modelling~\cite{Sand2016,Andersson2016}, but upscaling relies on
knowledge of dislocation density~\cite{Teng2000} or step
densities~\cite{Bracco2013}. AFM studies see only single screw
dislocations and not ensembles of them on entire crystal surfaces.

Batch experiments measure (and sometimes control) fluid supersaturation during growth of a large number of calcite grains that are present as crystallization seeds.~\cite{Reddy1981,Andreassen2004} The calculation of crystal growth rate from such experiments is highly dependent on the estimation of reactive surface area of the seed crystals. Batch measurements have the advantage of measuring average rates over thousands of surfaces but cannot observe the mechanisms of crystal growth.

Calcite {\em dissolution} rates measured by different methods show large variation~\cite{Arvidson2003,Luttge2013,Colombani2016}. In order to bridge the gap in information and scales of AFM measurements and batch measurements for calcite {\em dissolution}, whole crystal studies have been performed by interferometry ex situ~\cite{Arvidson2003} and in situ~\cite{Neuville2017}.  These {\em dissolution} studies of whole crystals do not rely on extra parameters like dislocation density or reactive surface area We claim that more whole crystal studies of CaCO$_3$ {\em growth} under well controlled conditions are also necessary and present here a novel microfluidic and microscopy method to fill the gap in length scales.

\paragraph{Microfluidics and CaCO$_3$\\}
Compared to batch methods and AFM, microfluidics represents a relatively young
but promising platform to run nucleation and growth experiments due to its high
level of control over diffusion, concentration, flow dynamics,
water-oil interface, liquid-gas interface and other factors to the
degree that bulk methods can hardly reach. 

\red{Most microfluidic studies dealing with CaCO$_3$ focus on nucleation and early growth of different CaCO$_3$ polymorphs. Some study hydrophobicity of CaCO$_3$ coatings~\cite{Pohl2014,Chen2012}, use CaCO$_3$ as source for Ca ions to produce alginate hydrogels or capsules~\cite{Liu2013,Shi2009,Amici2008,Tan2007,Huang2007} or to form hollow CaCO$_3$ nanoparticles for drug delivery~\cite{Xu2012}. Other studies of dissolution~\cite{Neuville2017}, reactive
flow~\cite{Zhang2010,Boyd2014,Pribyl2005} and calcite wetting of single or
multiphase flow~\cite{Song2014,Wang2017} have also been performed in
microfluidic networks.

Studies in nucleation and growth of CaCO$_3$ polymorphs in microfluidics are mainly intended to cast light on fundamental aspects of inorganic CaCO$_3$ crystallization pathways~\cite{Rodriguez-Ruiz2014,Li2017a,Zeng2018} or biomimetic or biomineralization processes~\cite{Yin2009,Neira-Carrillo2009,Neira-Carrillo2010,Yin2010,Ji2010,Yashina2012,Seo2013,Gong2015,Beuvier2015,Kim2017}. 

Zeng et al~\cite{Zeng2018} used microfluidics to control the formation of ACC and transformation pathways to vaterite and calcite, Li et al~\cite{Li2017a} also studied crystallization pathways but included the use of FTIR in microfluidics and Rodriguez-Ruiz~\cite{Rodriguez-Ruiz2014} studied how confinement in microfluidic reactors stabilized the hydrated CaCO$_3$ mineral ikaite. Yashina et al~\cite{Yashina2012} used ex situ SEM to study nucleation and initial growth of CaCO$_3$ polymorphs formed by mixing
Na$_2$CO$_3$ and CaCl$_2$ in microfluidic droplets, Gong et al~\cite{Gong2015}, used a gas-liquid microfluidic system to study growth of calcite
crystals around obstacles and Kim et al~\cite{Kim2017} used a similar system to
study early stages of CaCO$_3$ growth in presence of
additives. Seo et al~\cite{Seo2013} mixed Na$_2$CO$_3$
and CaCl$_2$ in a microfluidic linear gradient mixer to study CaCO$_3$
crystallization morphologies. In another set of
studies~\cite{Yin2009,Yin2010,Ji2010} they mixed Na$_2$CO$_3$ and
CaCl$_2$ with and without mollusk shell proteins in a microfluidic
T-junction to study CaCO$_3$ crystallization polymorphs. A similar
T-junction and fluids were employed to study CaCO$_3$ polymorphs in
situ by synchrotron X-ray methods~\cite{Beuvier2015}. Neira-Carrillo~\cite{Neira-Carrillo2009,Neira-Carrillo2010} studied templated CaCO$_3$ growth on different polymers. }

CaCO$_3$ nucleation studies usually involve segmented-flow mixing of
reagents, i.e., nucleation occurs within droplets, while very few
operate under continuous-flow conditions. The novelty of the
present study is the use of a microfluidic
continuous flow reactor to induce nucleation and slow growth of CaCO$_3$ crystals from the surrounding solution under highly controlled conditions. This also enables the study of thermodynamically unstable polymorphs.

\section{Experimental}
\begin{figure}[th]
\includegraphics[width=8cm]{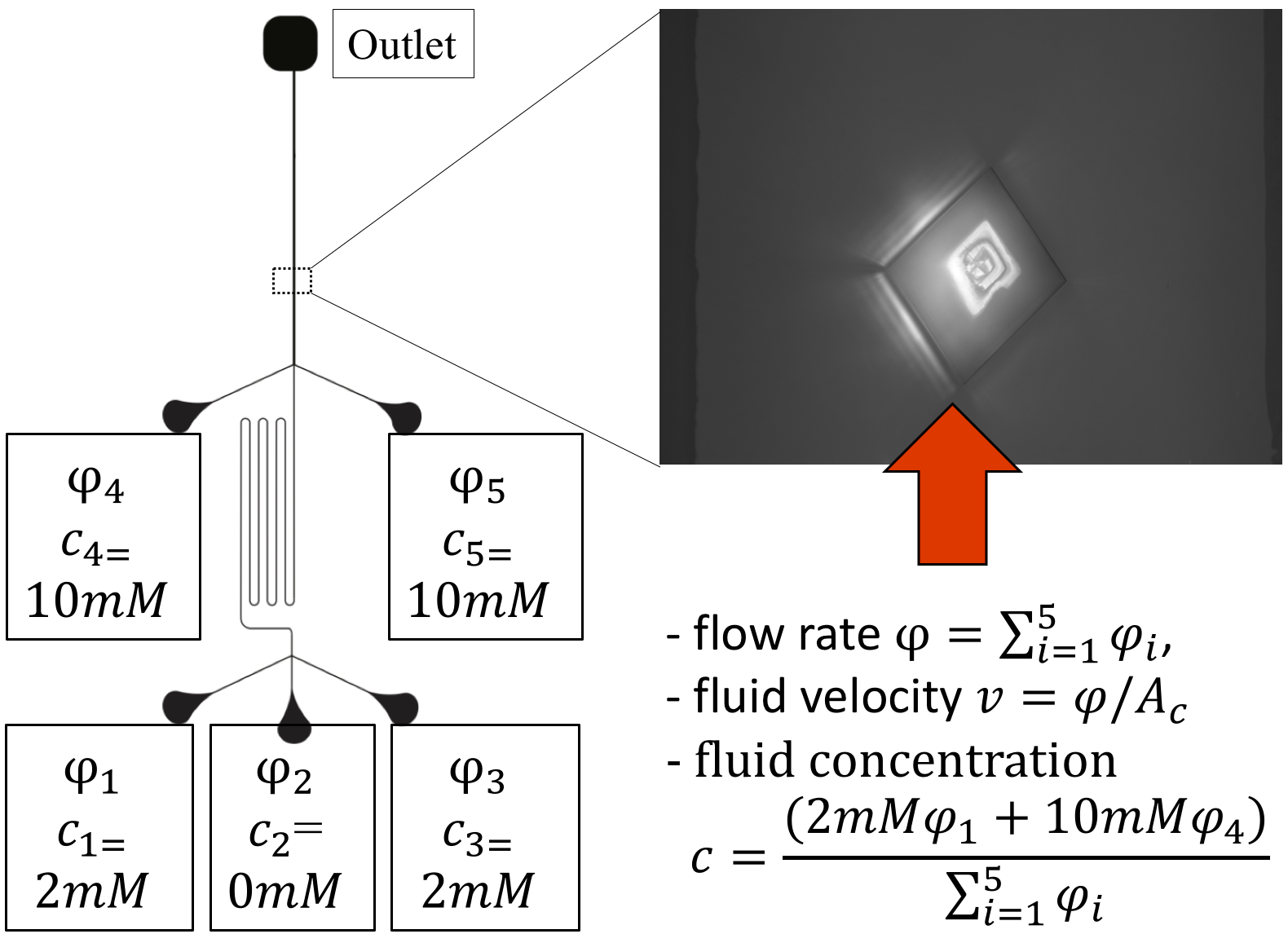}
\begin{tabular}{lcccccccc}\hline\hline          &$\varphi_1$&$\varphi_2$&$\varphi_3$&$\varphi_4$&$\varphi_5$&$\varphi$&$u$&$c$\\
           &\tiny$\frac{\mu{\rm l}}{{\rm min}}$
           &\tiny$\frac{\mu{\rm l}}{{\rm min}}$
           &\tiny$\frac{\mu{\rm l}}{{\rm min}}$
           &\tiny$\frac{\mu{\rm l}}{{\rm min}}$
           &\tiny$\frac{\mu{\rm l}}{{\rm min}}$
           &\tiny$\frac{\mu{\rm l}}{{\rm min}}$
           &\tiny$\frac{{\rm mm}}{{\rm s}}$&\tiny mM\\\hline
\scriptsize Nucl. & 1 & 0.5 & 1 & 2 & 2 & 6.5& 20 & 3.4\\
\scriptsize Growth     & 1 & 0.5 & 1 & 0 & 0 & 2.5 & 7.7 & 0.8\\
\scriptsize Sat.  & 1 & 2 & 1 & 0 & 0 & 4 & 12 & 0.5\\\hline\hline
\end{tabular}
\caption{\em\footnotesize Flow rate control for nucleation and growth. The fluid concentration in the channel at the crystal depends on the relative flow rates of the 5 inlets. CaCl$_2$ is injected through inlets 1 and 4 and Na$_2$CO$_3$ is injected through inlets 3 and 5.
Inlets 4 and 5 are only used during nucleation to assure nucleation in the channel between the second junction and the outlet. During growth the concentration is varied by changing $\varphi_2$ in the range shown in the table.}
\label{fig:flowgrowth}
\end{figure}

The aim of the experimental setup is to nucleate calcium carbonate crystals in a limited area that allows: i) high resolution imaging access, ii) controlling which polymorph to keep in the system for study, iii) controlling the saturation conditions at the growing crystal surface, iv) allowing slow growth of crystals from the nuclei, v) measuring growth rates, and vi) avoiding clogging of the microfluidic device due to crystal growth elsewhere in the system.

\subsection{Microfluidic devices}
\begin{figure}[ht]
\includegraphics[width=8cm]{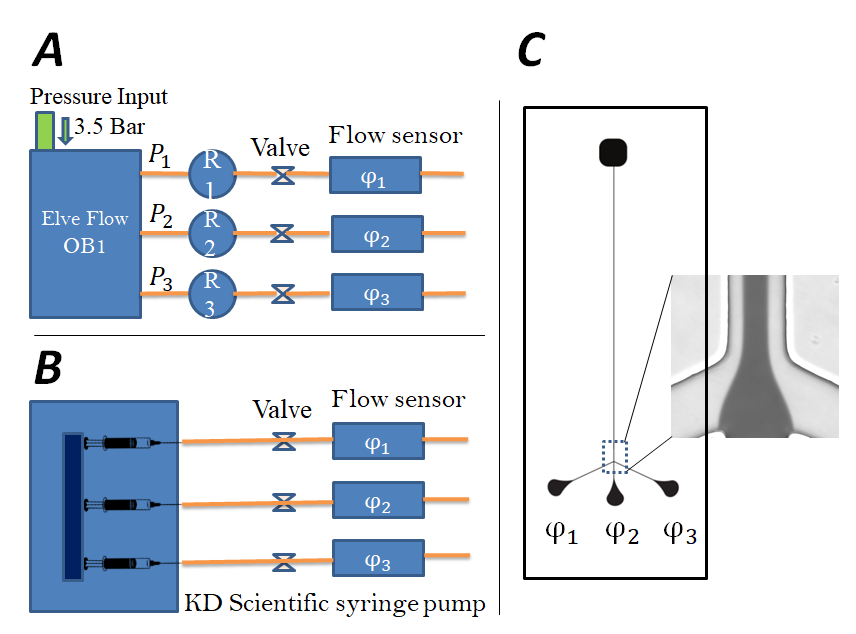}
\caption{\em\footnotesize Flow rate control for flow stability tests. The two system A and B were connected to the microfluidic channels with pure water in channels 1 and 3 and dyed water in channel 2. System A could operate in two feedback modes: pressure control or flow control. The syringe pump was run with either glass or plastic syringes. The inset image shows the junctions where the coloured and uncoloured flows meet. These images and recordings from the flow sensors were used as independent measurements of flow balance in the channel.}
\label{fig:flowstab_config}
\end{figure}

Two different microfluidic patterns were designed for this study. Figure~\ref{fig:flowgrowth} shows the pattern used for crystal nucleation and growth experiments where the channel dimensions are 120$\pm 2$~$\mu$m wide and 45~$\mu$m high, the length from first to second junction is $l_c$=50~mm and the length from the second junction to the outlet is 10~mm. Figure~\ref{fig:flowstab_config} shows the pattern used for flow stability experiments where the channel dimensions are 70$\pm 2$~$\mu$m wide and 45~$\mu$m high and the length from the junction to outlet is 20~mm.

The channel networks were designed in Adobe Illustrator, saved as pdf and used to print the photomask on a film substrate (Selba S.A, www.selba.ch). A  photoresist (SU-8 GM1070, Gersteltec, www.gersteltec.ch) was spun on silicon wafers at different speeds for different thicknesses, UV radiated (UV-KUB2, http://www.kloe.fr) and developed with PGMEA (www.sigmaaldrich.com) according to producers data sheet. Channel networks were cast in PDMS (Dow Corning Sylgard-184A, www.sigmaaldrich.com) with a 1:10 elastomer to curing agent ratio. Inlets and outlets of 1.5 mm diameter were punched subsequently and both  the PDMS and glass covers were treated with corona plasma (Electro-Technic Model BD-20V, http://www.electrotechnicproducts.com) before assembly.

\subsection{Nucleation and growth of carbonate crystals}
Pure water, CaCl$_2$ solution and  Na$_2$CO$_3$ solution are injected through separate inlets to the microfluidic device. The solutions meet at the junctions and the solutions are mixed by diffusion transversal to the flow. The fluid concentration in the channel at the crystal depends on the relative flow rates, $\phi_i$, at the 5 inlets that have fixed solution concentrations (see Figure~\ref{fig:flowgrowth}). Separate sections below will treat the difficulty of maintaining sufficiently good control of flow rates during the experiments. To ensure good mixing of the solution, the diffusion time scale for mixing, $\tau_D=(w/2)^2/D=3$~s (where $w$=120~$\mu$m is the channel width and $D=1.1\cdot 10^{-9}$~m/s is the mean diffusion coefficient of the ions in water), is smaller than the time, $\tau_{mix}=l_c/u=4-7$~s, it takes for the fluid to arrive at the growing crystal. 

Nucleation of a crystal within the microfluidic channel was carried out in two steps. First, the channel was filled with deionized water at a flow rate $\varphi_1=0.5\mu$l/min from inlet 2. 2mM CaCl$_2$ and Na$_2$CO$_3$ solutions were subsequently injected at inlets 1 and 3 to achieve a CaCO$_3$ concentration of $c$=0.8~mM. This value was sufficiently low to avoid any nucleation. When the flow reached a stable behaviour, the 10mM CaCl$_2$ and Na$_2$CO$_3$ solutions, were injected into the channel from inlets 4 and 5 (using 2.5mL Halmiton 1000 syringes on a KD Scientific Legato 180 syringe pump) to achieve a CaCO$_3$ concentration of $c=$3.4~mM. 

Once we observed crystals sticking to the glass or to the PDMS surface, the flows $\varphi_4$ and $\varphi_5$ were stopped. There is a certain probability that the first nucleus be either calcite, vaterite or aragonite, in our conditions the probability was roughly 50/50 calcite/vaterite in agreement with the measurements of Ogino et al~\cite{Ogino1987a}. In the remainder of the experiment we always chose to observe the crystal furthest upstream to be sure that the fluid concentration was determined by the flow rates and was not affected by other crystals upstream. Since the crystal is unaffected by other crystals and fully controlled by the concentration of the flowing solution we could dissolve and grow any CaCO$_3$ polymorph at will. Before calcite growth experiments we would dissolve vaterite at $c=0.5$~mM while keeping the calcite crystal unchanged. We performed calcite growth rate experiments with concentrations in the range 0.55-0.8~mM.

\begin{figure}[th]
\includegraphics[width=8cm]{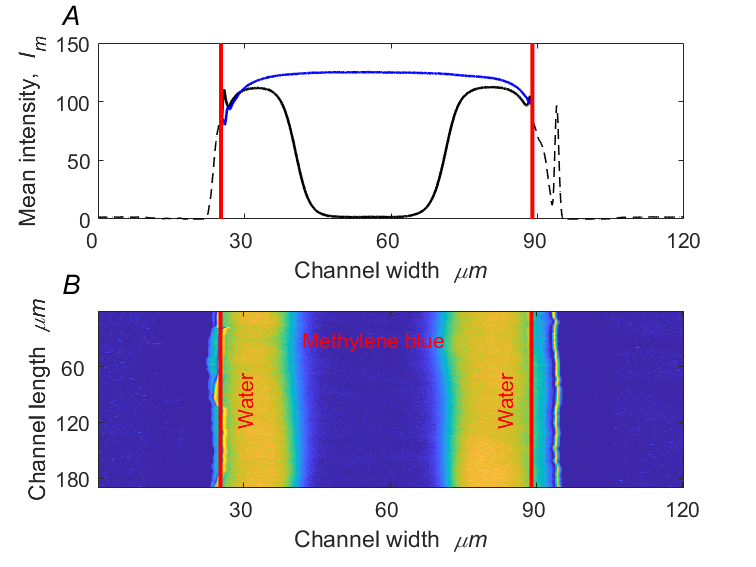}
\caption{\em\footnotesize Calculation of water and methylene-blue relative volume ratio from difference image. {\bf Bottom:} The intensity difference between an image of water and methylene blue flowing in a part of the PDMS channel. {\bf Top:} The mean intensities I$_m$ along the width of the channel for water/methylene-blue/water solution (black line) and only water in the channel (blue line). The red vertical lines define the edge of the channel, x$_1$ and x$_2$. The black dashed line show the mean intensity outside the channel.}
\label{fig:flowstab_calc}
\end{figure}

\subsection{Flow rate control}
In our microfluidic experiments, the flow stability is extremely important to achieve the stable concentrations necessary for crystal growth. After struggling with many random nucleation and dissolution events in our first setup using syringe pumps we decided to test the stability of the flows. The stability of the syringe pump and gas pressure control systems are separately tested as shown in Figure~\ref{fig:flowstab_config}. The gas pressure control system includes an Elveflow controller (Elveflow OB1 mk3, www.elveflow.com) which has both flow rate and pressure control modes, flow valves (Elveflow MUX) and flow sensors (0.4-7~$\mu$L/min, Elveflow). The input pressures $P_i$ are controlled by the OB1. The three inlets have the same flow resistance so that the pressure control mode is achieved by setting the same inlet pressures $P_1=P_2=P_3$.
In flow rate mode, OB1 adjusts the pressure to keep each flow rate constant. The syringe pump system includes a syringe pump (KD Scientific legato 180, www.kdscientific.com), BD plastic syringe (BD science, www.bd.com), Halmiton glass syringe (Halmiton 1000 syringe series, www.hamiltoncompany.com), flow valve (Elveflow MUX) and flow sensors (0.4-7~$\mu$L/min, Elveflow). The flow rate is controlled by the speed of the plunger and the inner diameter of the syringe. Because the accuracy of the syringe diameters of plastic and glass syringes differ, we test their stability separately.

Figure~\ref{fig:flowstab_config} displays the configurations used to test the flow stability: water was injected into the channel from inlets 1 and 3  with the flow rates $\varphi_1$ and $\varphi_3$ and methylene-blue solution (CAS Number: 61-73-4, Aldon Corp www.aldon-chem.com) was injected through inlet 2 at flow rate $\varphi_2$ which created two water/methylene-blue interfaces. Images of the flows (see Figures~\ref{fig:flowstab_config}C and \ref{fig:flowstab_calc}B) was followed for 5 hours. By averaging the intensity of the image along Y direction, we plotted the averaged intensity I$_m$ along X direction (Figure~\ref{fig:flowstab_calc}A). The channel edges x$_1$ and x$_2$ and the interface of the water/methylene blue were clearly identified by thresholding (see red lines in Figure~\ref{fig:flowstab_calc}A). The relative volume of water in the channel $\gamma$ was used to study the flow stability:
\begin{equation}
\gamma=\frac{\bar{I_m}}{\bar{I_w}}=\frac{\int_{x_1}^{x_2}I_m dx}{\int_{x_1}^{x_2}I_w dx},
\end{equation}
where $\bar{I_m}$ is the average image intensity of the water/methylene-blue/water solution in the channel (Figure~\ref{fig:flowstab_calc}A black line). And $\bar{I_w}$ means the average image intensity when there is only water in the channel (Figure~\ref{fig:flowstab_calc}A blue line) and $x_1$ and $x_2$ are the edges of the channel (Figure~\ref{fig:flowstab_calc}A red lines).

For the crystal growth experiments the flow rates $\varphi_1$, $\varphi_2$, $\varphi_3$, are CaCl$_2$, water, Na$_2$CO$_3$ instead of water, methylene-blue, water. The same flow rate fraction,$\gamma$, is proportional to the CaCO$_3$ concentration, $c$:
\begin{equation}
c=\frac{(c'\varphi_1+c'\varphi_3)/2}{\sum_i\varphi_i}=\frac{c'}{2}\frac{\bar{I_m}}{\bar{I_w}}=\frac{c'}{2}\gamma,
\label{eq:gamma}
\end{equation}
where $c'$ is the concentration of the CaCl$_2$ and Na$_2$CO$_3$ solutions and the stability of $\gamma$ shows the stability of the final CaCO$_3$ concentration $c$ during crystal growth.

\subsection{Solution preparation, concentration and saturation calculations}
Na$_2$CO$_3$ and CaCl$_2$ 10~mM stock solutions were prepared using a
balance (Mettler AE260 Delta Range) and deionized water (Millipore
Direct-Q 3UV) as solvent: 122.7 mg of Na$_2$CO$_3$ (VWR Ref.27767.364
Assay 99.0~100.5\% ) were dissolved in 115.6 mL water and 126.0 mg of
CaCl$_2·$2H$_2$O (VWR Ref.22322.364 Assay 97.0~103.0\% ) were
dissolved in 85.7 mL water. Then, they were left to equilibrate with
atmospheric CO$_2$ for 48 hours. Subsequently, 2 mM solutions were
prepared by diluting the stock solutions with deionized water (2mL
from the stock solution were mixed with 8mL of deionized water) prior
to each experiment. After dilution, they were immediately entered into
pressure flasks using air at pressures between 1 and 2.2 atmospheres
(absolute pressure).

The saturation index, $\Sigma$ has been calculated by
PHREEQC~\cite{Charlton2011}. The supersaturation is
$\Omega=IAP/K_{sp}$ and the saturation index is
\begin{equation}
\Sigma=\frac{\Delta\mu}{kT}=\ln(\frac{a_{Ca^{2+}}a_{CO_3^{2-}} }{K_{sp}})=\ln(\Omega)
\end{equation}
We have used the value of the solubility product $K_{sp}=10^{-8.54}$
that \citeauthor{Teng2000}~\cite{Teng2000} found to correspond to when
spirals on the 10$\bar{1}$4 surface stopped growing. They used
slightly different fluids than in our study and a fixed pH of 8.5
whereas our solutions were not buffered and the pH varied. The lack of
constraints on pH causes some inaccuracy of the calculation of
saturation index. The calcite crystals in this study changed from
growth to dissolution at a concentration of $c_{sat}$=0.5~mM. The
calculated saturation index at this concentration is $\Sigma$=-0.11.
We assume that the inaccuracy is mainly a shift of the saturation
point and that the saturation index has the correct dependency on
concentration.

An unintended side effect of using air as pressurizing gas for flow
control was that it allowed CO$_2$ at higher pressures to exchange
with the solution. The final supersaturation of the fluid is
influenced by the amount of absorbed CO$_2$. The pressure corrected
saturation index $\Sigma_p$ of the solutions is calculated for
solutions in equilibrium with air at the actual pressure driving the
flow.  Since the exchange between CO$_2$ in the air and the solution
takes time and the gas pressure changes during an experiment, the
equilibrium CO$_2$ concentration in the fluid at the driving pressure
is an upper bound.

\subsection{Imaging}
All experiments were monitored using an Olympus GX71 microscope with a
green LED light source with a wavelength of 550 nm (from ThorLabs
www.thorlabs.com). UPLanFLN 100x/1.30 and UPLanFI 40x/0.75p objectives
with resolution $d$=260~nm and 450~nm (from Olympus
www.olympus-lifescience.com). Images were recorded using a Pointgrey
camera (Mono, Grasshopper3, GS3-U3-91S6M-C, www.ptgrey.com) with 3376
x 2704 resolution and saved as 8 bit TIFF files. For the high
resolution objective there were 5 pixels per resolution diameter $d$.
The analysis of the image sequences was carried out using in-house
developed scripts in Matlab (www.mathworks.com) and ImageJ
(imagej.nih.gov/ij/).

\subsection{Determination of growth rate}
\begin{figure}[th]
\includegraphics[width=8cm]{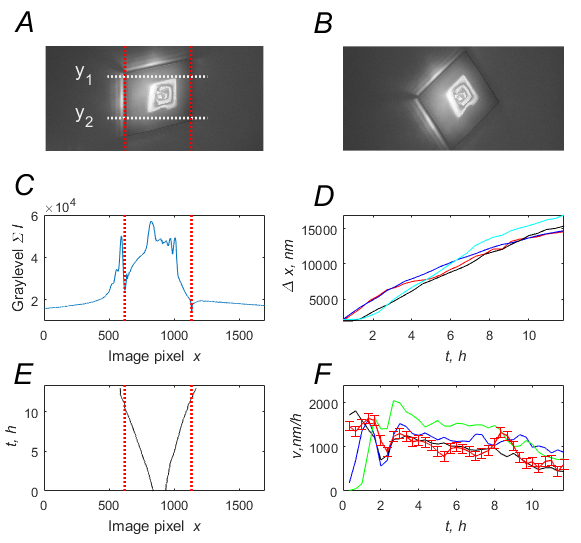}
\caption{\em\footnotesize {\bf Calculation of growth rates.} Original
  image of crystal in {\bf B} is rotated to have two faces vertical in
  {\bf A} and the intensities between $y_1$ and $y_2$ are summed to
  produce the grey level plot in {\bf C}: $\sum_{i=1}^2I(y_i)$. The
  local minima in the grey level identify the crystal face indicated by
  the red vertical lines. The minima positions in the images as
  function of time are displayed in {\bf E} and the change in
  position, $\Delta x$, in {\bf D}. Linear fits to $\Delta x(t)$ in
  smaller time windows yield the growth rates $v$ in {\bf F}.}
\label{fig:radius_measure}
\end{figure}
Calcite is a trigonal-rhombohedral crystal and one ($10\bar{1}4$) face
is at rest and parallel with the glass surface (see
Figure~\ref{fig:radius_measure}). We measure the imaged position of
the four crystal ($10\bar{1}4$) faces that are almost perpendicular to
the glass surface neglecting the difference between the calcite
surface normal and the projection along the imaging plane. The
positions of the four faces are determined by averaging intensities
along the direction parallel to the face and finding the minimum
intensity. The averaging along the crystal faces allows determination
of the crystal face position relative to earlier positions with a
resolution of $\pm 1$~pixel~$=\pm 50$~nm.  The growth rates are
determined by linear fits to the position data as function of time and
the reported precision of the growth rates are the standard deviations
of these fits.

\section{Results and discussion}
\subsection{Flow stability}
\begin{figure}[th]
\includegraphics[width=8cm]{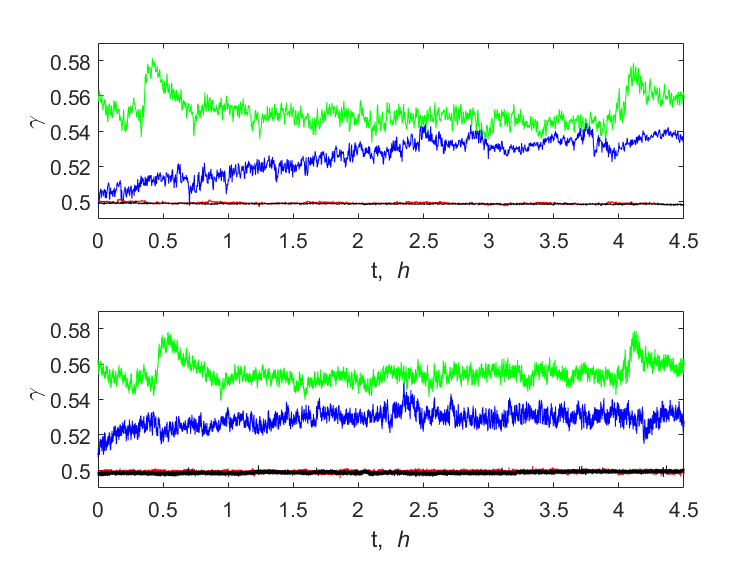}
\begin{tabular}{ccccc} 
\hline\hline
    &  \multicolumn{2}{c}{ $100\cdot\sigma_\gamma/\bar{\gamma}$}
    & \multicolumn{2}{c}{ $100\cdot\Delta_\gamma/\bar{\gamma}$} \\
    & Image & FS & Image & FS \\\hline
  GS  & 1.4 &1.0& 8.8 &7.1\\
  PS  &  2.0 &0.9& 9.0  &8.1\\
  FC  &0.1 &0.1& 0.8 &1.4\\
  PC  & 0.1 &0.1&0.7  &1.3\\\hline
\hline
\end{tabular}
\caption{\em\footnotesize Stability of water/dyed water/water flow. The relative volume $\gamma$ of dyed water calculated from images (top) and flow sensors (bottom) is displayed in {\bf black:} pressure control ({\bf PC}), {\bf red:} flow control ({\bf FC}), {\bf green:} plastic syringe ({\bf PS}) and {\bf blue:} glass syringe ({\bf GS}). Table: the relative standard deviation $\sigma_\gamma/\bar{\gamma}$ and the relative maximum deviation $\Delta_\gamma/\bar{\gamma}$ of the four flow stability tests.}
\label{fig:flowstab}
\end{figure}
The stability of the CaCO$_3$ concentration, $c$, which is proportional to the volume fraction in the flow (see equation (\ref{eq:gamma})) is key to accurate and reliable measurements of crystal growth rates. The flow stability tests were performed for 5 hours with 2 water input channels at 1~$\mu$l/min and 1 input channel with dyed water at 2~$\mu$l/min and in four configurations: Gas pressure driven fluid flow with 1) pressure control ({\bf PC}) and 2) flow control ({\bf FC}) and syringe pump with 3) plastic syringe ({\bf PS}) and 4) glass syringe ({\bf GS}). The instantaneous volume fraction of dyed water presented in Figure~\ref{fig:flowstab} was calculated from images according to equation (\ref{eq:gamma}) and from the flow sensors.

\begin{figure}[th]
\includegraphics[width=8cm]{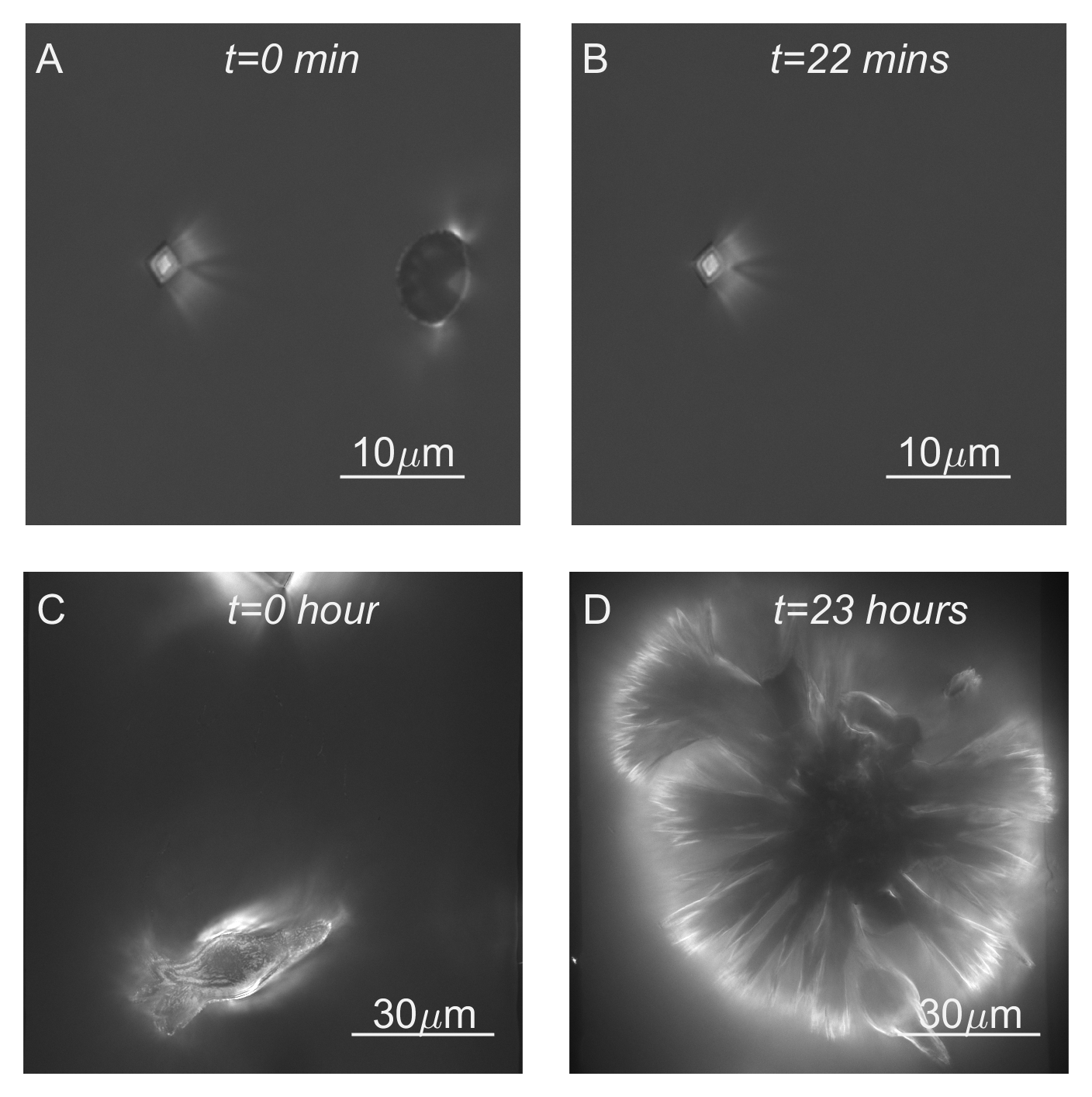}
\caption{\em\footnotesize {\bf Top (A and B):} A. Calcite and Vaterite nuclei at the same time and attached at the middle of the channel. B. The vaterite disappeared after 22 minutes flushing with 0.5mM CaCO$_3$ solution. {\bf Bottom C.D:} C. A aragonite located in the channel. D. The aragonite grows to fill the channel after 7 hours growth at 0.8 mM CaCO$_3$ concentration. The radius of aragonite crystal increase at a rate of 4160 nm/h.}
\label{fig:nucleation}
\end{figure}

In Figure~\ref{fig:flowstab} one observes that the flow control by syringe pump causes both constant deviation from the desired flow fraction, $\gamma=0.5$, slow drift and large, sudden changes compared to the gas pressure driven flows. The constant deviation and slow drift are probably caused by differences in inner diameter of the syringes or by mechanical inaccuracies of the syringe pump. The short term fluctuations are probably caused by stick slip of the drive train in the pump or/and by the plunger in the syringe.
We have also characterized the flow stability by the relative standard deviation of $\gamma$, $\sigma_\gamma/\bar{\gamma}$ and the relative difference between the maximum and minimum value, $\Delta_\gamma/\bar{\gamma}$ (where $\bar{\gamma}$ is the average of $\gamma$) displayed in Figure~\ref{fig:flowstab}. These measures demonstrate that the fluctuations in flow fraction (and thereby concentration) is one order of magnitude larger for the syringe pump. The relative maximum deviation, $\Delta_\gamma/\bar{\gamma}$, captures rare events of high and low concentrations and is related to the probability of uncontrolled nucleation events during the crystal growth. The large fluctuations for the syringe pump explains our problems of sudden nucleation events and ruined experiments while using syringe pumps. The gas pressure flow control system on the other hand provides a stable concentration during all our many hours long experiments with a standard deviation of only 0.1\% and maximum deviation of only 1\%. The pressure control (PC) mode shows the best performance but cannot be used for growth experiment since the crystal growth increases the pressure in the channel with time. All the crystal growth experiments reported here are therefore performed using the flow control (FC) mode.

\subsection{Calcite nucleation}
At saturation index $\Sigma=1.9$, nucleation occurs fast and some nuclei attach either to the PDMS or glass surfaces.  Figure~\ref{fig:nucleation} A shows both a calcite and a vaterite nuclei present simultaneously in the middle of the channel. Since  mixing occurs by diffusion, the highest concentration appears at the centre of the channel and the nuclei attach at the middle of the glass or PDMS surface allowing for good optical access and free space for subsequent growth. Changing the calcium carbonate concentration to 0.5~mM, the vaterite is totally dissolved after 22 minutes and the calcite is left unchanged (Figure~\ref{fig:nucleation} B). The  calcite nuclei are predominantly rhombohedral and the 10$\bar{1}$4 surface attach to the glass or PDMS. The crystals can then grow freely on 5 of the 6 surfaces. Figure~\ref{fig:nucleation} C shows a vaterite nucleus that is chosen for growth in the channel and Figure~\ref{fig:nucleation} D shows the same crystal after 23 hours growth. This demonstrates how the microfluidic device can keep any polymorph stable for further study.

\subsection{Calcite growth rate}
\begin{figure*}
\includegraphics[width=16cm]{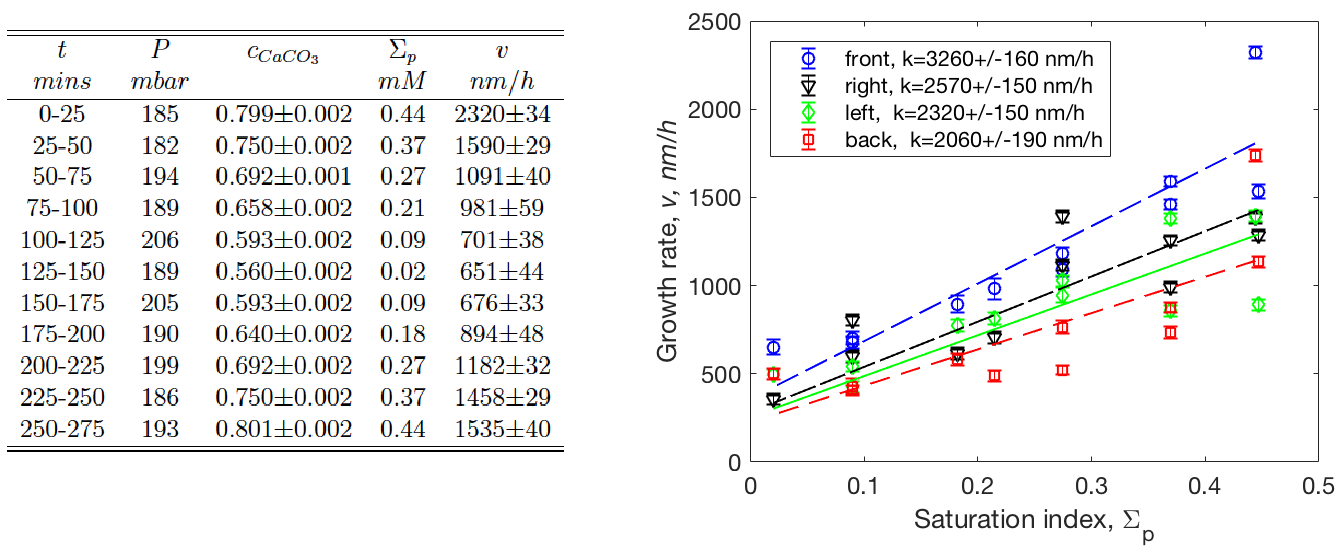} 
 \caption{\em\footnotesize Growth conditions and rates of crystal
   G. In less than 5 hours growth rates of four crystal faces were
   measured twice at 6 different CaCO$_3$ concentrations
   $c_{CaCO_3}$. Saturation indices $\Sigma_p$ are calculated at
   CaCO$_3$ concentrations $c$ and CO$_2$ partial pressures
   [mbar] $P_{CO_2}=3.9\cdot 10^{-4}(1013+P)$, where $P$ is the
   pressure (in this table) to drive the fluid flow. The
   growth rates in the table are those of the front face which is
   almost perpendicular to the fluid flow, the figure shows growth
   rates of all faces. The slopes of the
   fitted lines are the growth rate constants, $k$. The growth rate
   constant at the crystal surface facing the fluid flow is 60\%
   higher than that of the surface at the back of the crystal due to
   the increasing effect of the diffusion boundary layer with distance
   of flow along the crystal. }
\label{fig:CrystG}
\end{figure*}
We have performed very many growth rate experiments with the setup
reported here. We repeatedly measured two limits to growth: Below
$c=0.5$~mM calcite crystals started dissolving, above this
concentration they were growing and above $c=0.8$~mM there
was a risk of nucleation of new crystals in the device.
Figure~\ref{fig:CrystG} shows the results of an experiment with fast
change of solution concentration to obtain growth rates over the whole
range of concentrations in practice available to this device. The
experiment was performed with decreasing and then increasing
concentrations in fast succession to check the ``stability'' of the
growth (see results and discussion of growth rate dispersion /
variability below). The growth rates, $v$, were determined from images from
periods of 15 of the 25 minutes at each concentration. For the lowest
growth rates of about 400~nm/h this means that the change in position
of the crystal face was only 100~nm during 15 minutes. The error bars
in Figure~\ref{fig:CrystG} are the standard deviations of the linear
fits to the 60 positions recorded during the 15 minute period. These
results show that the technique is fast and accurate.

\red{ Teng et al~\cite{Teng2000} summarized macroscopic growth rate laws as belonging to two types:
\begin{eqnarray}
v&=&k'\left(e^{n\Sigma}-1\right)\approx k'n\Sigma \approx k\Sigma\label{eq:k}, {\hspace{0.2cm}\rm and}\\
v&=&k_2\left(e^{\Sigma}-1\right)^n\approx k_2\Sigma^n\label{eq:k2},
\end{eqnarray}
where the leading order approximations are valid when $\Sigma\ll 1$. Teng et al~\cite{Teng2000} found that the surface processes (growth only at screw dislocations) they observed and the rate they measured at $\Sigma< 1.5$ was consistent with equation~(\ref{eq:k2}) with $n=2$. The growth rates we have measured are more consistent with equation~(\ref{eq:k}) in the small range of
concentration and saturation indices. The growth rate constants $k$ and $k_2$
are a good basis for comparison with other data from the
literature.} The growth rate constants determined here and displayed in
Figure~\ref{fig:CrystG} have a standard deviation of 5-9\%.

The results in Figure~\ref{fig:CrystG} are for a crystal with two
faces perpendicular and two faces parallel to the channel walls. The
front face that receives fluid unperturbed by solid surfaces or
crystal growth grows 30\% faster than the two faces parallel to the
fluid flow and 60\% faster than the back surface facing
downstream. This can be explained by the hydrodynamic conditions at
the crystal surfaces. The fluid velocity, $u$, is zero at the surfaces
and grows towards the middle of the channel. This results in a
hydrodynamic boundary layer where only diffusion transports ions to
the surface. The thickness, $\delta$, of this boundary layer scales
with the fluid velocity, $u$, and length of travel along the crystal
surface, $l$, as $\delta\propto\sqrt{l/u}$. Calculating the boundary
layer thickness on the back surface we find the ratio between the
observed, diffusion limited growth rate, $v_d$, and the purely
reaction limited growth rate, $v_r$, (see the Supplementary
information) $v_d/v_r\sim 0.4$. This agrees well with the observed
difference between the front and the back of crystal G of about
60\%. Since the growth rates of the front surface are the closest we
can come to pure reaction limited growth rate this is what we report
in the table in Figure~\ref{fig:CrystG}.

\begin{table}
  \begin{tabular}{cccccc}
    \hline\hline
    Exp & $P$    &$c$ &$\Sigma_p$& $v$ & $\delta_v$\\
        & $mbar$ &$mM$&          & $nm/h$& $nm/h$\\\hline
    A     & 200  & 0.499 & -0.11 & 8    & 10 \\
    B     & 400  & 0.706 & 0.24  & 1030 & 40 \\
    C     & 540  & 0.718 & 0.21  & 1020 & 60 \\
    D     & 250  & 0.79  & 0.41  & 1580 & 100 \\
    E     & 360  & 0.801 & 0.4   & 1600 & 100 \\
    F     & 200  & 0.801 & 0.45  & 1010 & 50 \\
%    A$_2$ & 18-27    & 170  & 0.619 & 0.15  & 197  & 10 \\
%    C$_2$ & 7.5-11.5 & 1050 & 0.707 & 0.079 & 60   & 50 \\
%    C$_3$ &11.5-16.5 & 720  & 0.716 & 0.17  & 400  & 50 \\
    \hline
    \hline
  \end{tabular}
\caption{\em\footnotesize Growth conditions and growth rates of
  crystals A-F. Saturation indices $\Sigma_p$ are calculated at
  CaCO$_3$ concentrations $c_{CaCO_3}$ and CO$_2$
  partial pressures [mbar] $P_{CO_2}=3.9\cdot
  10^{-4}(1013+P)$, where $P$ is the pressure (in this
  table) to drive the fluid flow. }
  \label{tab:growthrates}
\end{table}

For most other crystals we have performed 5-20 hours growth
experiments at the same concentration in the range 0.5-0.8~mM (see
Table~\ref{tab:growthrates}). In some cases the pressure increased
during the experiment (due to crystal growth at the outlet) and we
discard that part of the experiment since we could not measure the
CO$_2$ concentration of the solution. In Table~\ref{tab:growthrates}
we report the growth rate of the crystal surfaces growing fastest
since they were facing the flow and less perturbed by the hydrodynamic
boundary layer.

\begin{figure}[th]
\includegraphics[width=9cm]{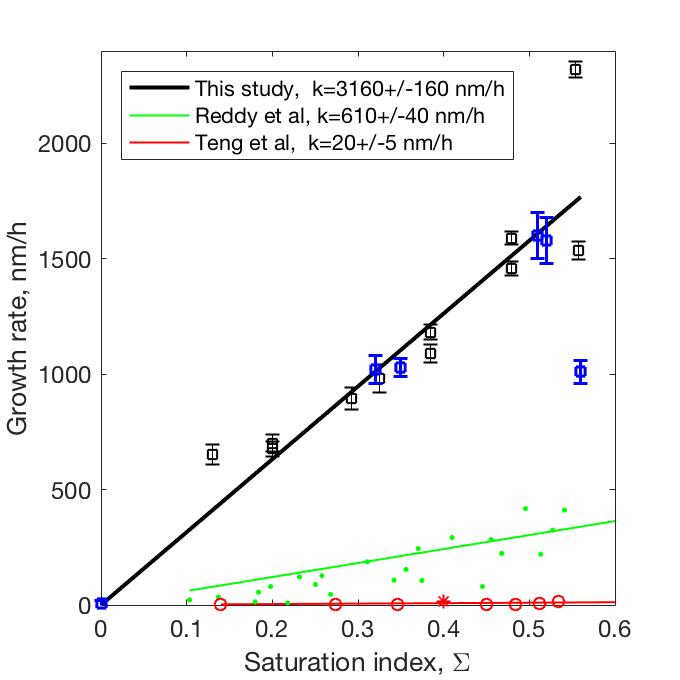}
\caption{\em\footnotesize {\bf Comparison of growth rate constants.} Blue
  squares: $v_{max}$ from Table~\ref{tab:growthrates}, black squares:
  $v$ from table in Figure~\ref{fig:CrystG}, green dots: growth rates from
  Figure~3 of Reddy et al.~\cite{Reddy1981}, red circles: growth rates
from Table~2 of Teng et al.\cite{Teng2000}, red asterisk: growth rate
from Bracco et al\cite{Bracco2013}. The saturation index of our measurements have been corrected to be zero when there is no growth: $\Sigma=\Sigma_p+0.11$.
}
\label{fig:growthrate_SI}
\end{figure}

In Figure~\ref{fig:growthrate_SI} we have plotted our measured growth
rates versus the calculated saturation indices (see also
Figure~\ref{fig:CrystG}). We have added 0.11 to the values $\Sigma_p$
in Table~\ref{tab:growthrates} to account for the observed saturation
concentration $c_{sat}=0.5$~mM (at 25$^\circ$C and $p=200~mbar$) of
our solutions. We have fit equation (\ref{eq:k}) to
the growth rates in Table~\ref{tab:growthrates} and obtain the growth
constant $k=3260\pm 160$~nm/h. \red{At $\Sigma< 0.7$ we have
used equation (\ref{eq:k}) to fit to calcite growth data measured by
AFM~\cite{Teng2000,Bracco2013}, $k_{AFM}=20\pm 5$~nm/h, and by batch
crystallization~\cite{Reddy1981}, $k_{Reddy}=610\pm 40$~nm/h. To our
great surprise our growth rate constant is about 5 times higher than
$k_{Reddy}$ and about 160 times higher than $k_{AFM}$. If we use the growth rate equation~(\ref{eq:k2}) instead with a wider range of $\Sigma$, our growth rate constant $k_2$ is a factor 95 larger than that from Teng et al.}
We have checked and rechecked our lab notebooks and all our calculations several
times. The experiments have been performed over a period of 3 years
and in different physical versions of the same layout. The solutions
have been prepared from scratch several times, the flow rates have
been verified by many methods and calculations, the imaging length
scale has been recalibrated and the imaging frame rate has been
rechecked and verified.

\red{Our first explanation for the two orders of magnitude difference was that the AFM
experiments are much more affected by hydrodynamic boundary layers
than our experiments. Traditional analysis of AFM flow cells indicate that this
should not account for the large discrepancy: A typical AFM flow
cell~\cite{Schmidt1994} has $l\sim 10^{-3}$~m and $u\sim
10^{-2}$~m/s. Since the flow is over a flat crystal surface $s_c/s\sim
1$. Although the boundary layer is 10 times thicker, the final
reduction in observed growth rate in AFM experiments is then expected
to be $v_d/v_r\sim 0.2$, only a factor 2 smaller than the back side of
the crystals in our experiments. This means that in such a case we
expect no more than a factor 2 difference between the growth rate on
our front face and those in AFM measurements. However, a recent study by Peruffo et al~\cite{Peruffo2016} has critically analyzed the effect of dissolution/growth outside the region of interest (ROI) of the AFM measurement and the effect of spatially heterogeneous flow in AFM flow cells. They found that in the case of gypsum this could cause 1-2 orders of magnitude lower dissolution rates measured by AFM than by batch experiments.}

Another difference between our measurements and AFM measurements is
that our calcite crystals are untouched by human hands and have never
experienced any other environment than the growth solution, whereas
AFM experiments are typically performed on flat portions of cleaved
Island spar crystals. It is conceivable that either contaminations in
the crystal or strain energy due to the cleaving make these crystals
less reactive than our newly nucleated calcite rhombs. \red{Many
  practitioners do, however etch the cleaved calcite surface before
  starting growth in order to avoid this problem.} It is possible that
Island spar crystals that have grown very slowly over geological time
have a very small dislocation density compared to our
crystals. However, from a geometrical consideration, the fastest
growth of a whole crystal will occur when the crystal face is covered
by a single growth spiral around a dislocation core. \red{If
  hydrodynamics and processes outside the ROI of the AFM cannot
  account for the discrepancy with our measurements then we must
  consider either rough step fronts instead of strait step trains or a
  much higher step density.}

It is outside the scope of this paper to explain why growth rate constants for Island spar calcite crystals calculated from AFM measurements are 2 orders of magnitude smaller than we measure. Our measurements should, however prompt a serious discussion of what are actually the rate limiting mechanisms in different calcite growth rate experiments.

\section{Conclusions}
A novel method for studying nucleation and growth of CaCO$_3$ crystals
in situ has been developed and tested rigorously. We demonstrate that
precise flow control is essential and how this is achieved. The method
has the advantage that one may study single crystals of polymorphs
that are thermodynamically unstable in collections of many crystals
and that one obtains precise and accurate growth rates without any
extra assumptions. We also demonstrate that at low supersaturations
where 2D nucleation does not occur we measure the growth rate constant
of calcite to be \red{5 times} larger than that reported by batch methods
(that need additional measurement or assumption of reactive surface
area) and \red{two orders of magnitude} larger than measured by AFM.  Considering the
large interest in calcite growth in for example geoscience,
environmental science, and industry we consider that it is important
to explain the discrepancy of growth rate constants between different
methods.  \red{The method presented here can easily be applied to many other minerals.}

\begin{acknowledgement}
This project has received funding from the European Union's Horizon
2020 research and innovation programme under the Marie
Sklodowska-Curie grant agreement no.\ 642976 (ITN NanoHeal) and from
the Norwegian Research Council grant no.\ 222386.

\end{acknowledgement}

\begin{suppinfo}

Additional information on experimental details and calculation of
hydrodynamic boundary layers are available.

\end{suppinfo}

\bibliography{calcite_growth}

\providecommand{\latin}[1]{#1}
\makeatletter
\providecommand{\doi}
  {\begingroup\let\do\@makeother\dospecials
  \catcode`\{=1 \catcode`\}=2 \doi@aux}
\providecommand{\doi@aux}[1]{\endgroup\texttt{#1}}
\makeatother
\providecommand*\mcitethebibliography{\thebibliography}
\csname @ifundefined\endcsname{endmcitethebibliography}
  {\let\endmcitethebibliography\endthebibliography}{}
\begin{mcitethebibliography}{54}
\providecommand*\natexlab[1]{#1}
\providecommand*\mciteSetBstSublistMode[1]{}
\providecommand*\mciteSetBstMaxWidthForm[2]{}
\providecommand*\mciteBstWouldAddEndPuncttrue
  {\def\EndOfBibitem{\unskip.}}
\providecommand*\mciteBstWouldAddEndPunctfalse
  {\let\EndOfBibitem\relax}
\providecommand*\mciteSetBstMidEndSepPunct[3]{}
\providecommand*\mciteSetBstSublistLabelBeginEnd[3]{}
\providecommand*\EndOfBibitem{}
\mciteSetBstSublistMode{f}
\mciteSetBstMaxWidthForm{subitem}{(\alph{mcitesubitemcount})}
\mciteSetBstSublistLabelBeginEnd
  {\mcitemaxwidthsubitemform\space}
  {\relax}
  {\relax}

\bibitem[Banner(1995)]{Banner1995}
Banner,~J.~L. {Application of the trace element and isotope geochemistry of
  strontium to studies of carbonate diagenesis}. \emph{Sedimentology}
  \textbf{1995}, \emph{42}, 805--824\relax
\mciteBstWouldAddEndPuncttrue
\mciteSetBstMidEndSepPunct{\mcitedefaultmidpunct}
{\mcitedefaultendpunct}{\mcitedefaultseppunct}\relax
\EndOfBibitem
\bibitem[Riding(2000)]{Riding2000}
Riding,~R. {Microbial carbonates: the geological record of calcified
  bacterial-algal mats and biofilms}. \emph{Sedimentology} \textbf{2000},
  \emph{47}, 179--214\relax
\mciteBstWouldAddEndPuncttrue
\mciteSetBstMidEndSepPunct{\mcitedefaultmidpunct}
{\mcitedefaultendpunct}{\mcitedefaultseppunct}\relax
\EndOfBibitem
\bibitem[Fairchild and Treble(2009)Fairchild, and Treble]{Fairchild2009}
Fairchild,~I.~J.; Treble,~P.~C. {Trace elements in speleothems as recorders of
  environmental change}. \emph{Quaternary Science Reviews} \textbf{2009},
  \emph{28}, 449--468\relax
\mciteBstWouldAddEndPuncttrue
\mciteSetBstMidEndSepPunct{\mcitedefaultmidpunct}
{\mcitedefaultendpunct}{\mcitedefaultseppunct}\relax
\EndOfBibitem
\bibitem[Roehl and Choquette(1985)Roehl, and Choquette]{Roehl1985}
Roehl,~P.~O.; Choquette,~P.~W. \emph{Carbonate Petroleum Reservoirs};
  1985\relax
\mciteBstWouldAddEndPuncttrue
\mciteSetBstMidEndSepPunct{\mcitedefaultmidpunct}
{\mcitedefaultendpunct}{\mcitedefaultseppunct}\relax
\EndOfBibitem
\bibitem[Watabe(1981)]{Watabe1981}
Watabe,~N. {Crystal Growth of Calcium Carbonate in the Invertibrates}.
  \emph{Progress in crystal growth and characterization of materials}
  \textbf{1981}, \emph{4}, 99--147\relax
\mciteBstWouldAddEndPuncttrue
\mciteSetBstMidEndSepPunct{\mcitedefaultmidpunct}
{\mcitedefaultendpunct}{\mcitedefaultseppunct}\relax
\EndOfBibitem
\bibitem[Colfen(2003)]{Colfen2003}
Colfen,~H. {Precipitation of carbonates: recent progress in controlled
  production of complex shapes}. \emph{Current opinion in colloid and interface
  science} \textbf{2003}, \emph{8}, 23--31\relax
\mciteBstWouldAddEndPuncttrue
\mciteSetBstMidEndSepPunct{\mcitedefaultmidpunct}
{\mcitedefaultendpunct}{\mcitedefaultseppunct}\relax
\EndOfBibitem
\bibitem[Meldrum(2003)]{Meldrum2003}
Meldrum,~F.~C. {Calcium carbonate in biomineralisation and biomimetic
  chemistry}. \emph{IInternational Materials Review} \textbf{2003}, \emph{48},
  187--224\relax
\mciteBstWouldAddEndPuncttrue
\mciteSetBstMidEndSepPunct{\mcitedefaultmidpunct}
{\mcitedefaultendpunct}{\mcitedefaultseppunct}\relax
\EndOfBibitem
\bibitem[Carr and Frederick(2000)Carr, and Frederick]{Carr2000}
Carr,~F.~P.; Frederick,~D.~K. {Calcium carbonate}. 2000\relax
\mciteBstWouldAddEndPuncttrue
\mciteSetBstMidEndSepPunct{\mcitedefaultmidpunct}
{\mcitedefaultendpunct}{\mcitedefaultseppunct}\relax
\EndOfBibitem
\bibitem[Bracco \latin{et~al.}(2013)Bracco, Stack, and Steefel]{Bracco2013}
Bracco,~J.~N.; Stack,~A.~G.; Steefel,~C.~I. {Upscaling calcite growth rates
  from the mesoscale to the macroscale}. \emph{Environmental Science and
  Technology} \textbf{2013}, \emph{47}, 7555--7562\relax
\mciteBstWouldAddEndPuncttrue
\mciteSetBstMidEndSepPunct{\mcitedefaultmidpunct}
{\mcitedefaultendpunct}{\mcitedefaultseppunct}\relax
\EndOfBibitem
\bibitem[Andersson \latin{et~al.}(2016)Andersson, Dobbersch{\"{u}}tz, Sand,
  Tobler, {De Yoreo}, and Stipp]{Andersson2016}
Andersson,~M.~P.; Dobbersch{\"{u}}tz,~S.; Sand,~K.~K.; Tobler,~D.~J.; {De
  Yoreo},~J.~J.; Stipp,~S.~L. {A Microkinetic Model of Calcite Step Growth}.
  \emph{Angewandte Chemie - International Edition} \textbf{2016}, \emph{55},
  11086--11090\relax
\mciteBstWouldAddEndPuncttrue
\mciteSetBstMidEndSepPunct{\mcitedefaultmidpunct}
{\mcitedefaultendpunct}{\mcitedefaultseppunct}\relax
\EndOfBibitem
\bibitem[Reddy \latin{et~al.}(1981)Reddy, Plummer, and Busenberg]{Reddy1981}
Reddy,~M.~M.; Plummer,~L.~N.; Busenberg,~E. {Crystal growth of calcite from
  calcium bicarbonate solutions at constant PCO2and 25°C: a test of a calcite
  dissolution model}. \emph{Geochimica et Cosmochimica Acta} \textbf{1981},
  \emph{45}, 1281--1289\relax
\mciteBstWouldAddEndPuncttrue
\mciteSetBstMidEndSepPunct{\mcitedefaultmidpunct}
{\mcitedefaultendpunct}{\mcitedefaultseppunct}\relax
\EndOfBibitem
\bibitem[Hillner \latin{et~al.}(1992)Hillner, Manne, Gratz, and
  Hansma]{Hillner1992}
Hillner,~P.~E.; Manne,~S.; Gratz,~A.~J.; Hansma,~P.~K. {AFM images of
  dissolution and growth on a calcite crystal}. \emph{Ultramicroscopy}
  \textbf{1992}, \emph{42-44}, 1387--1393\relax
\mciteBstWouldAddEndPuncttrue
\mciteSetBstMidEndSepPunct{\mcitedefaultmidpunct}
{\mcitedefaultendpunct}{\mcitedefaultseppunct}\relax
\EndOfBibitem
\bibitem[Stipp \latin{et~al.}(1994)Stipp, Eggleston, and Nielsen]{Stipp1994}
Stipp,~S. L.~S.; Eggleston,~C.~M.; Nielsen,~B.~S. {Calcite Surface-Structure
  Observed at Microtopographic and Molecular Scales with Atomic-Force
  Microscopy (AFM)}. \emph{Geochimica Et Cosmochimica Acta} \textbf{1994},
  \emph{58}, 3023--3033\relax
\mciteBstWouldAddEndPuncttrue
\mciteSetBstMidEndSepPunct{\mcitedefaultmidpunct}
{\mcitedefaultendpunct}{\mcitedefaultseppunct}\relax
\EndOfBibitem
\bibitem[Teng(1998)]{Teng1998}
Teng,~H.~H. {Thermodynamics of calcite growth: baseline for understanding
  biomineral formation}. \emph{Science} \textbf{1998}, \emph{282},
  724--727\relax
\mciteBstWouldAddEndPuncttrue
\mciteSetBstMidEndSepPunct{\mcitedefaultmidpunct}
{\mcitedefaultendpunct}{\mcitedefaultseppunct}\relax
\EndOfBibitem
\bibitem[Teng \latin{et~al.}(1999)Teng, Dove, and Deyoreo]{Teng1999}
Teng,~H.~H.; Dove,~P.~M.; Deyoreo,~J.~J. {Reversed calcite morphologies induced
  by microscopic growth kinetics: Insight into biomineralization}.
  \emph{Geochimica et Cosmochimica Acta} \textbf{1999}, \emph{63},
  2507--2512\relax
\mciteBstWouldAddEndPuncttrue
\mciteSetBstMidEndSepPunct{\mcitedefaultmidpunct}
{\mcitedefaultendpunct}{\mcitedefaultseppunct}\relax
\EndOfBibitem
\bibitem[Teng \latin{et~al.}(2000)Teng, Dove, and {De Yoreo}]{Teng2000}
Teng,~H.~H.; Dove,~P.~M.; {De Yoreo},~J.~J. {Kinetics of calcite growth:
  Surface processes and relationships to macroscopic rate laws}.
  \emph{Geochimica et Cosmochimica Acta} \textbf{2000}, \emph{64},
  2255--2266\relax
\mciteBstWouldAddEndPuncttrue
\mciteSetBstMidEndSepPunct{\mcitedefaultmidpunct}
{\mcitedefaultendpunct}{\mcitedefaultseppunct}\relax
\EndOfBibitem
\bibitem[Ruiz-Agudo and Putnis(2012)Ruiz-Agudo, and Putnis]{Ruiz-Agudo2012}
Ruiz-Agudo,~E.; Putnis,~C.~V. {Direct observations of mineral fluid reactions
  using atomic force microscopy: the specific example of calcite}.
  \emph{Mineralogical Magazine} \textbf{2012}, \emph{76}, 227--253\relax
\mciteBstWouldAddEndPuncttrue
\mciteSetBstMidEndSepPunct{\mcitedefaultmidpunct}
{\mcitedefaultendpunct}{\mcitedefaultseppunct}\relax
\EndOfBibitem
\bibitem[Sand \latin{et~al.}(2016)Sand, Tobler, Dobbersch{\"{u}}tz, Larsen,
  Makovicky, Andersson, Wolthers, and Stipp]{Sand2016}
Sand,~K.~K.; Tobler,~D.~J.; Dobbersch{\"{u}}tz,~S.; Larsen,~K.~K.;
  Makovicky,~E.; Andersson,~M.~P.; Wolthers,~M.; Stipp,~S.~L. {Calcite Growth
  Kinetics: Dependence on Saturation Index, Ca2+:CO32-Activity Ratio, and
  Surface Atomic Structure}. \emph{Crystal Growth and Design} \textbf{2016},
  \emph{16}, 3602--3612\relax
\mciteBstWouldAddEndPuncttrue
\mciteSetBstMidEndSepPunct{\mcitedefaultmidpunct}
{\mcitedefaultendpunct}{\mcitedefaultseppunct}\relax
\EndOfBibitem
\bibitem[Andreassen and Hounslow(2004)Andreassen, and Hounslow]{Andreassen2004}
Andreassen,~J.~P.; Hounslow,~M.~J. {Growth and aggregation of vaterite in
  seeded-batch experiments}. \emph{AIChE Journal} \textbf{2004}, \emph{50},
  2772--2782\relax
\mciteBstWouldAddEndPuncttrue
\mciteSetBstMidEndSepPunct{\mcitedefaultmidpunct}
{\mcitedefaultendpunct}{\mcitedefaultseppunct}\relax
\EndOfBibitem
\bibitem[Arvidson \latin{et~al.}(2003)Arvidson, Ertan, Amonette, and
  Luttge]{Arvidson2003}
Arvidson,~R.~S.; Ertan,~I.~E.; Amonette,~J.~E.; Luttge,~A. {Variation in
  calcite dissolution rates: A fundamental problem?} \emph{Geochimica et
  Cosmochimica Acta} \textbf{2003}, \emph{67}, 1623--1634\relax
\mciteBstWouldAddEndPuncttrue
\mciteSetBstMidEndSepPunct{\mcitedefaultmidpunct}
{\mcitedefaultendpunct}{\mcitedefaultseppunct}\relax
\EndOfBibitem
\bibitem[L{\"{u}}ttge \latin{et~al.}(2013)L{\"{u}}ttge, Arvidson, and
  Fischer]{Luttge2013}
L{\"{u}}ttge,~A.; Arvidson,~R.~S.; Fischer,~C. {A stochastic treatment of
  crystal dissolution kinetics}. \emph{Elements} \textbf{2013}, \emph{9},
  183--188\relax
\mciteBstWouldAddEndPuncttrue
\mciteSetBstMidEndSepPunct{\mcitedefaultmidpunct}
{\mcitedefaultendpunct}{\mcitedefaultseppunct}\relax
\EndOfBibitem
\bibitem[Colombani(2016)]{Colombani2016}
Colombani,~J. {The Alkaline Dissolution Rate of Calcite}. \emph{Journal of
  Physical Chemistry Letters} \textbf{2016}, \emph{7}, 2376--2380\relax
\mciteBstWouldAddEndPuncttrue
\mciteSetBstMidEndSepPunct{\mcitedefaultmidpunct}
{\mcitedefaultendpunct}{\mcitedefaultseppunct}\relax
\EndOfBibitem
\bibitem[Neuville \latin{et~al.}(2017)Neuville, Renaud, Luu, Minde, Jettestuen,
  Vinningland, Hiorth, and Dysthe]{Neuville2017}
Neuville,~A.; Renaud,~L.; Luu,~T.~T.; Minde,~M.~W.; Jettestuen,~E.;
  Vinningland,~J.~L.; Hiorth,~A.; Dysthe,~D.~K. {Xurography for microfluidics
  on a reactive solid}. \emph{Lab Chip} \textbf{2017}, \emph{17}, 293\relax
\mciteBstWouldAddEndPuncttrue
\mciteSetBstMidEndSepPunct{\mcitedefaultmidpunct}
{\mcitedefaultendpunct}{\mcitedefaultseppunct}\relax
\EndOfBibitem
\bibitem[Pohl and Weiss(2014)Pohl, and Weiss]{Pohl2014}
Pohl,~A.; Weiss,~I.~M. {Real-time monitoring of calcium carbonate and cationic
  peptide deposition on carboxylate-SAM using a microfluidic SAW biosensor}.
  \emph{Beilstein Journal of Nanotechnology} \textbf{2014}, \emph{5},
  1823--1835\relax
\mciteBstWouldAddEndPuncttrue
\mciteSetBstMidEndSepPunct{\mcitedefaultmidpunct}
{\mcitedefaultendpunct}{\mcitedefaultseppunct}\relax
\EndOfBibitem
\bibitem[Chen \latin{et~al.}(2012)Chen, Wan, and Xu]{Chen2012}
Chen,~P.-C.; Wan,~L.-S.; Xu,~Z.-K. {Bio-inspired CaCO3 coating for
  superhydrophilic hybrid membranes with high water permeability}.
  \emph{Journal of Materials Chemistry} \textbf{2012}, \emph{22}, 22727\relax
\mciteBstWouldAddEndPuncttrue
\mciteSetBstMidEndSepPunct{\mcitedefaultmidpunct}
{\mcitedefaultendpunct}{\mcitedefaultseppunct}\relax
\EndOfBibitem
\bibitem[Liu \latin{et~al.}(2013)Liu, Wu, Ju, Xie, Wang, Niu, and Chu]{Liu2013}
Liu,~L.; Wu,~F.; Ju,~X.~J.; Xie,~R.; Wang,~W.; Niu,~C.~H.; Chu,~L.~Y.
  {Preparation of monodisperse calcium alginate microcapsules via internal
  gelation in microfluidic-generated double emulsions}. \emph{Journal of
  Colloid and Interface Science} \textbf{2013}, \emph{404}, 85--90\relax
\mciteBstWouldAddEndPuncttrue
\mciteSetBstMidEndSepPunct{\mcitedefaultmidpunct}
{\mcitedefaultendpunct}{\mcitedefaultseppunct}\relax
\EndOfBibitem
\bibitem[Shi \latin{et~al.}(2009)Shi, Tsao, Yang, Liu, Dykstra, Rubloff,
  Ghodssi, Bentley, and Payne]{Shi2009}
Shi,~X.~W.; Tsao,~C.~Y.; Yang,~X.; Liu,~Y.; Dykstra,~P.; Rubloff,~G.~W.;
  Ghodssi,~R.; Bentley,~W.~E.; Payne,~G.~F. {Electroaddressing of cell
  populations by co-deposition with calcium alginate hydrogels}. \emph{Advanced
  Functional Materials} \textbf{2009}, \emph{19}, 2074--2080\relax
\mciteBstWouldAddEndPuncttrue
\mciteSetBstMidEndSepPunct{\mcitedefaultmidpunct}
{\mcitedefaultendpunct}{\mcitedefaultseppunct}\relax
\EndOfBibitem
\bibitem[Amici \latin{et~al.}(2008)Amici, Tetradis-Meris, de~Torres, and
  Jousse]{Amici2008}
Amici,~E.; Tetradis-Meris,~G.; de~Torres,~C.~P.; Jousse,~F. {Alginate gelation
  in microfluidic channels}. \emph{Food Hydrocolloids} \textbf{2008},
  \emph{22}, 97--104\relax
\mciteBstWouldAddEndPuncttrue
\mciteSetBstMidEndSepPunct{\mcitedefaultmidpunct}
{\mcitedefaultendpunct}{\mcitedefaultseppunct}\relax
\EndOfBibitem
\bibitem[Tan and Takeuchi(2007)Tan, and Takeuchi]{Tan2007}
Tan,~W.~H.; Takeuchi,~S. {Monodisperse alginate hydrogel microbeads for cell
  encapsulation}. \emph{Advanced Materials} \textbf{2007}, \emph{19},
  2696--2701\relax
\mciteBstWouldAddEndPuncttrue
\mciteSetBstMidEndSepPunct{\mcitedefaultmidpunct}
{\mcitedefaultendpunct}{\mcitedefaultseppunct}\relax
\EndOfBibitem
\bibitem[Huang(2007)]{Huang2007}
Huang,~K.-S. {Using a microfluidic chip and internal gelation reaction for
  monodisperse calcium alginate microparticles generation}. \emph{Frontiers in
  Bioscience} \textbf{2007}, \emph{12}, 3061\relax
\mciteBstWouldAddEndPuncttrue
\mciteSetBstMidEndSepPunct{\mcitedefaultmidpunct}
{\mcitedefaultendpunct}{\mcitedefaultseppunct}\relax
\EndOfBibitem
\bibitem[Xu \latin{et~al.}(2012)Xu, Yang, Xu, Xia, and Chen]{Xu2012}
Xu,~B.-Y.; Yang,~Z.-Q.; Xu,~J.-J.; Xia,~X.-H.; Chen,~H.-Y. {Liquid–gas dual
  phase microfluidic system for biocompatible CaCO3 hollow nanoparticles
  generation and simultaneous molecule doping}. \emph{Chemical Communications}
  \textbf{2012}, \emph{48}, 11635--11637\relax
\mciteBstWouldAddEndPuncttrue
\mciteSetBstMidEndSepPunct{\mcitedefaultmidpunct}
{\mcitedefaultendpunct}{\mcitedefaultseppunct}\relax
\EndOfBibitem
\bibitem[Zhang \latin{et~al.}(2010)Zhang, Dehoff, Hess, Oostrom, Wietsma,
  Valocchi, Fouke, and Werth]{Zhang2010}
Zhang,~C.; Dehoff,~K.; Hess,~N.; Oostrom,~M.; Wietsma,~T.~W.; Valocchi,~A.~J.;
  Fouke,~B.~W.; Werth,~C.~J. {Pore-scale study of transverse mixing induced
  CaCO3 precipitation and permeability reduction in a model subsurface
  sedimentary system}. \emph{Environmental Science and Technology}
  \textbf{2010}, \emph{44}, 7833--7838\relax
\mciteBstWouldAddEndPuncttrue
\mciteSetBstMidEndSepPunct{\mcitedefaultmidpunct}
{\mcitedefaultendpunct}{\mcitedefaultseppunct}\relax
\EndOfBibitem
\bibitem[Boyd \latin{et~al.}(2014)Boyd, Yoon, Zhang, Oostrom, Hess, Fouke,
  Valocchi, and Werth]{Boyd2014}
Boyd,~V.; Yoon,~H.; Zhang,~C.; Oostrom,~M.; Hess,~N.; Fouke,~B.;
  Valocchi,~A.~J.; Werth,~C.~J. {Influence of Mg2+ on CaCO3 precipitation
  during subsurface reactive transport in a homogeneous silicon-etched pore
  network}. \emph{Geochimica et Cosmochimica Acta} \textbf{2014}, \emph{135},
  321--335\relax
\mciteBstWouldAddEndPuncttrue
\mciteSetBstMidEndSepPunct{\mcitedefaultmidpunct}
{\mcitedefaultendpunct}{\mcitedefaultseppunct}\relax
\EndOfBibitem
\bibitem[Přibyl \latin{et~al.}(2005)Přibyl, {\v{S}}nita, and
  Marek]{Pribyl2005}
Přibyl,~M.; {\v{S}}nita,~D.; Marek,~M. {Nonlinear phenomena and qualitative
  evaluation of risk of clogging in a capillary microreactor under imposed
  electric field}. \emph{Chemical Engineering Journal} \textbf{2005},
  \emph{105}, 99--109\relax
\mciteBstWouldAddEndPuncttrue
\mciteSetBstMidEndSepPunct{\mcitedefaultmidpunct}
{\mcitedefaultendpunct}{\mcitedefaultseppunct}\relax
\EndOfBibitem
\bibitem[Song \latin{et~al.}(2014)Song, de~Haas, Fadaei, and Sinton]{Song2014}
Song,~W.; de~Haas,~T.~W.; Fadaei,~H.; Sinton,~D. {Chip-off-the-old-rock: the
  study of reservoir-relevant geological processes with real-rock micromodels}.
  \emph{Lab Chip} \textbf{2014}, \emph{14}, 4382--4390\relax
\mciteBstWouldAddEndPuncttrue
\mciteSetBstMidEndSepPunct{\mcitedefaultmidpunct}
{\mcitedefaultendpunct}{\mcitedefaultseppunct}\relax
\EndOfBibitem
\bibitem[Wang \latin{et~al.}(2017)Wang, Chang, and Gizzatov]{Wang2017}
Wang,~W.; Chang,~S.; Gizzatov,~A. {Toward Reservoir-on-a-Chip: Fabricating
  Reservoir Micromodels by in Situ Growing Calcium Carbonate Nanocrystals in
  Microfluidic Channels}. \emph{ACS Applied Materials and Interfaces}
  \textbf{2017}, \emph{9}, 29380--29386\relax
\mciteBstWouldAddEndPuncttrue
\mciteSetBstMidEndSepPunct{\mcitedefaultmidpunct}
{\mcitedefaultendpunct}{\mcitedefaultseppunct}\relax
\EndOfBibitem
\bibitem[Rodr{\'{i}}guez-Ruiz \latin{et~al.}(2014)Rodr{\'{i}}guez-Ruiz,
  Veesler, G{\'{o}}mez-Morales, Delgado-L{\'{o}}pez, Grauby, Hammadi, Candoni,
  and Garc{\'{i}}a-Ruiz]{Rodriguez-Ruiz2014}
Rodr{\'{i}}guez-Ruiz,~I.; Veesler,~S.; G{\'{o}}mez-Morales,~J.;
  Delgado-L{\'{o}}pez,~J.~M.; Grauby,~O.; Hammadi,~Z.; Candoni,~N.;
  Garc{\'{i}}a-Ruiz,~J.~M. {Transient calcium carbonate hexahydrate (ikaite)
  nucleated and stabilized in confined nano- and picovolumes}. \emph{Crystal
  Growth and Design} \textbf{2014}, \emph{14}, 792--802\relax
\mciteBstWouldAddEndPuncttrue
\mciteSetBstMidEndSepPunct{\mcitedefaultmidpunct}
{\mcitedefaultendpunct}{\mcitedefaultseppunct}\relax
\EndOfBibitem
\bibitem[Li \latin{et~al.}(2017)Li, Ihli, Marchant, Zeng, Chen, Wehbe, Cinque,
  Cespedes, Kapur, and Meldrum]{Li2017a}
Li,~S.; Ihli,~J.; Marchant,~W.~J.; Zeng,~M.; Chen,~L.; Wehbe,~K.; Cinque,~G.;
  Cespedes,~O.; Kapur,~N.; Meldrum,~F.~C. {Synchrotron FTIR mapping of
  mineralization in a microfluidic device}. \emph{Lab Chip} \textbf{2017},
  \emph{17}, 1616--1624\relax
\mciteBstWouldAddEndPuncttrue
\mciteSetBstMidEndSepPunct{\mcitedefaultmidpunct}
{\mcitedefaultendpunct}{\mcitedefaultseppunct}\relax
\EndOfBibitem
\bibitem[Zeng \latin{et~al.}(2018)Zeng, Cao, Wang, Guo, and Lu]{Zeng2018}
Zeng,~Y.; Cao,~J.; Wang,~Z.; Guo,~J.; Lu,~J. {Formation of Amorphous Calcium
  Carbonate and Its Transformation Mechanism to Crystalline CaCO
  {\textless}sub{\textgreater}3{\textless}/sub{\textgreater} in Laminar
  Microfluidics}. \emph{Crystal Growth {\&} Design} \textbf{2018}, \emph{18},
  1710--1721\relax
\mciteBstWouldAddEndPuncttrue
\mciteSetBstMidEndSepPunct{\mcitedefaultmidpunct}
{\mcitedefaultendpunct}{\mcitedefaultseppunct}\relax
\EndOfBibitem
\bibitem[Yin \latin{et~al.}(2009)Yin, Ji, Dobson, Mosbahi, Glidle, Gadegaard,
  Freer, Cooper, and Cusack]{Yin2009}
Yin,~H.; Ji,~B.; Dobson,~P.~S.; Mosbahi,~K.; Glidle,~A.; Gadegaard,~N.;
  Freer,~A.; Cooper,~J.~M.; Cusack,~M. {Screening of Biomineralization Using
  Microfluidics}. \emph{Analytical chemistry} \textbf{2009}, \emph{81},
  473--478\relax
\mciteBstWouldAddEndPuncttrue
\mciteSetBstMidEndSepPunct{\mcitedefaultmidpunct}
{\mcitedefaultendpunct}{\mcitedefaultseppunct}\relax
\EndOfBibitem
\bibitem[Neira-Carrillo \latin{et~al.}(2009)Neira-Carrillo, Pai,
  Fern{\'{a}}ndez, Carre{\~{n}}o, Quitral, and Arias]{Neira-Carrillo2009}
Neira-Carrillo,~A.; Pai,~R.~K.; Fern{\'{a}}ndez,~M.~S.; Carre{\~{n}}o,~E.;
  Quitral,~P.~V.; Arias,~J.~L. {Synthesis and characterization of sulfonated
  polymethylsiloxane polymer as template for crystal growth of CaCO3}.
  \emph{Colloid and Polymer Science} \textbf{2009}, \emph{287}, 385--393\relax
\mciteBstWouldAddEndPuncttrue
\mciteSetBstMidEndSepPunct{\mcitedefaultmidpunct}
{\mcitedefaultendpunct}{\mcitedefaultseppunct}\relax
\EndOfBibitem
\bibitem[Neira-Carrillo \latin{et~al.}(2010)Neira-Carrillo,
  V{\'{a}}squez-Quitral, Yazdani-Pedram, and Arias]{Neira-Carrillo2010}
Neira-Carrillo,~A.; V{\'{a}}squez-Quitral,~P.; Yazdani-Pedram,~M.; Arias,~J.~L.
  {Crystal growth of CaCO3induced by monomethylitaconate grafted
  polymethylsiloxane}. \emph{European Polymer Journal} \textbf{2010},
  \emph{46}, 1184--1193\relax
\mciteBstWouldAddEndPuncttrue
\mciteSetBstMidEndSepPunct{\mcitedefaultmidpunct}
{\mcitedefaultendpunct}{\mcitedefaultseppunct}\relax
\EndOfBibitem
\bibitem[Yin \latin{et~al.}(2010)Yin, Ji, Cusack, Freer, Dobson, Gadegaard, and
  Jiang]{Yin2010}
Yin,~H.; Ji,~B.; Cusack,~M.; Freer,~A.; Dobson,~P.~S.; Gadegaard,~N.; Jiang,~J.
  {Microfluidics in Biomineralization and Biomimicking Synthesis}. uTAS2010.
  2010; pp 986--988\relax
\mciteBstWouldAddEndPuncttrue
\mciteSetBstMidEndSepPunct{\mcitedefaultmidpunct}
{\mcitedefaultendpunct}{\mcitedefaultseppunct}\relax
\EndOfBibitem
\bibitem[Ji \latin{et~al.}(2010)Ji, Cusack, Freer, Dobson, Gadegaard, and
  Yin]{Ji2010}
Ji,~B.; Cusack,~M.; Freer,~A.; Dobson,~P.~S.; Gadegaard,~N.; Yin,~H. {Control
  of crystal polymorph in microfluidics using molluscan 28 kDa Ca2+-binding
  protein}. \emph{Integrative Biology} \textbf{2010}, \emph{2}, 528\relax
\mciteBstWouldAddEndPuncttrue
\mciteSetBstMidEndSepPunct{\mcitedefaultmidpunct}
{\mcitedefaultendpunct}{\mcitedefaultseppunct}\relax
\EndOfBibitem
\bibitem[Yashina \latin{et~al.}(2012)Yashina, Meldrum, and
  Demello]{Yashina2012}
Yashina,~A.; Meldrum,~F.; Demello,~A. {Calcium carbonate polymorph control
  using droplet-based microfluidics.} \emph{Biomicrofluidics} \textbf{2012},
  \emph{6}, 22001--2200110\relax
\mciteBstWouldAddEndPuncttrue
\mciteSetBstMidEndSepPunct{\mcitedefaultmidpunct}
{\mcitedefaultendpunct}{\mcitedefaultseppunct}\relax
\EndOfBibitem
\bibitem[Seo \latin{et~al.}(2013)Seo, Ko, Lee, and Kim]{Seo2013}
Seo,~S.~W.; Ko,~K.~Y.; Lee,~C.~S.; Kim,~I.~H. {CaCO3 Biomineralization in
  Microfluidic Crystallizer}. \emph{Korean Chem. Eng. Res.} \textbf{2013},
  \emph{51}, 151--156\relax
\mciteBstWouldAddEndPuncttrue
\mciteSetBstMidEndSepPunct{\mcitedefaultmidpunct}
{\mcitedefaultendpunct}{\mcitedefaultseppunct}\relax
\EndOfBibitem
\bibitem[Gong \latin{et~al.}(2015)Gong, Wang, Ihli, Kim, Li, Walshaw, Chen, and
  Meldrum]{Gong2015}
Gong,~X.; Wang,~Y.~W.; Ihli,~J.; Kim,~Y.~Y.; Li,~S.; Walshaw,~R.; Chen,~L.;
  Meldrum,~F.~C. {The Crystal Hotel: A Microfluidic Approach to Biomimetic
  Crystallization}. \emph{Advanced Materials} \textbf{2015}, \emph{27},
  7395--7400\relax
\mciteBstWouldAddEndPuncttrue
\mciteSetBstMidEndSepPunct{\mcitedefaultmidpunct}
{\mcitedefaultendpunct}{\mcitedefaultseppunct}\relax
\EndOfBibitem
\bibitem[Beuvier \latin{et~al.}(2015)Beuvier, Panduro, Kwa{\'{s}}niewski,
  Marre, Lecoutre, Garrabos, Aymonier, Calvignac, and Gibaud]{Beuvier2015}
Beuvier,~T.; Panduro,~E. A.~C.; Kwa{\'{s}}niewski,~P.; Marre,~S.; Lecoutre,~C.;
  Garrabos,~Y.; Aymonier,~C.; Calvignac,~B.; Gibaud,~A. {Implementation of in
  situ SAXS/WAXS characterization into silicon/glass microreactors}. \emph{Lab
  Chip} \textbf{2015}, \emph{15}, 2002--2008\relax
\mciteBstWouldAddEndPuncttrue
\mciteSetBstMidEndSepPunct{\mcitedefaultmidpunct}
{\mcitedefaultendpunct}{\mcitedefaultseppunct}\relax
\EndOfBibitem
\bibitem[Kim \latin{et~al.}(2017)Kim, Freeman, Gong, Levenstein, Wang, Kulak,
  Anduix-Canto, Lee, Li, Chen, Christenson, and Meldrum]{Kim2017}
Kim,~Y.-Y.; Freeman,~C.~L.; Gong,~X.; Levenstein,~M.~A.; Wang,~Y.; Kulak,~A.;
  Anduix-Canto,~C.; Lee,~P.~A.; Li,~S.; Chen,~L.; Christenson,~H.~K.;
  Meldrum,~F.~C. {The Effect of Additives on the Early Stages of Growth of
  Calcite Single Crystals}. \emph{Angewandte Chemie International Edition}
  \textbf{2017}, \emph{200444}, 11885--11890\relax
\mciteBstWouldAddEndPuncttrue
\mciteSetBstMidEndSepPunct{\mcitedefaultmidpunct}
{\mcitedefaultendpunct}{\mcitedefaultseppunct}\relax
\EndOfBibitem
\bibitem[Ogino \latin{et~al.}(1987)Ogino, Suzuki, and Sawada]{Ogino1987a}
Ogino,~T.; Suzuki,~T.; Sawada,~K. {The formation and transformation mechanism
  of calcium carbonate in water}. \emph{Geochimica et Cosmochimica Acta}
  \textbf{1987}, \emph{51}, 2757--2767\relax
\mciteBstWouldAddEndPuncttrue
\mciteSetBstMidEndSepPunct{\mcitedefaultmidpunct}
{\mcitedefaultendpunct}{\mcitedefaultseppunct}\relax
\EndOfBibitem
\bibitem[Charlton and Parkhurst(2011)Charlton, and Parkhurst]{Charlton2011}
Charlton,~S.~R.; Parkhurst,~D.~L. {Modules based on the geochemical model
  PHREEQC for use in scripting and programming languages}. \emph{Computers {\&}
  Geosciences} \textbf{2011}, \emph{37}, 1653--1663\relax
\mciteBstWouldAddEndPuncttrue
\mciteSetBstMidEndSepPunct{\mcitedefaultmidpunct}
{\mcitedefaultendpunct}{\mcitedefaultseppunct}\relax
\EndOfBibitem
\bibitem[Schmidt(1994)]{Schmidt1994}
Schmidt,~W.~U. {Use of Atomic Force Microscopy to Image Surfaces during Fluid
  Flow}. \emph{Journal of The Electrochemical Society} \textbf{1994},
  \emph{141}, L85\relax
\mciteBstWouldAddEndPuncttrue
\mciteSetBstMidEndSepPunct{\mcitedefaultmidpunct}
{\mcitedefaultendpunct}{\mcitedefaultseppunct}\relax
\EndOfBibitem
\bibitem[Peruffo \latin{et~al.}(2016)Peruffo, Mbogoro, Adobes-Vidal, and
  Unwin]{Peruffo2016}
Peruffo,~M.; Mbogoro,~M.~M.; Adobes-Vidal,~M.; Unwin,~P.~R. {Importance of Mass
  Transport and Spatially Heterogeneous Flux Processes for in Situ Atomic Force
  Microscopy Measurements of Crystal Growth and Dissolution Kinetics}.
  \emph{Journal of Physical Chemistry C} \textbf{2016}, \emph{120},
  12100--12112\relax
\mciteBstWouldAddEndPuncttrue
\mciteSetBstMidEndSepPunct{\mcitedefaultmidpunct}
{\mcitedefaultendpunct}{\mcitedefaultseppunct}\relax
\EndOfBibitem
\end{mcitethebibliography}


\providecommand{\latin}[1]{#1}
\makeatletter
\providecommand{\doi}
  {\begingroup\let\do\@makeother\dospecials
  \catcode`\{=1 \catcode`\}=2 \doi@aux}
\providecommand{\doi@aux}[1]{\endgroup\texttt{#1}}
\makeatother
\providecommand*\mcitethebibliography{\thebibliography}
\csname @ifundefined\endcsname{endmcitethebibliography}
  {\let\endmcitethebibliography\endthebibliography}{}
\begin{mcitethebibliography}{3}
\providecommand*\natexlab[1]{#1}
\providecommand*\mciteSetBstSublistMode[1]{}
\providecommand*\mciteSetBstMaxWidthForm[2]{}
\providecommand*\mciteBstWouldAddEndPuncttrue
  {\def\EndOfBibitem{\unskip.}}
\providecommand*\mciteBstWouldAddEndPunctfalse
  {\let\EndOfBibitem\relax}
\providecommand*\mciteSetBstMidEndSepPunct[3]{}
\providecommand*\mciteSetBstSublistLabelBeginEnd[3]{}
\providecommand*\EndOfBibitem{}
\mciteSetBstSublistMode{f}
\mciteSetBstMaxWidthForm{subitem}{(\alph{mcitesubitemcount})}
\mciteSetBstSublistLabelBeginEnd
  {\mcitemaxwidthsubitemform\space}
  {\relax}
  {\relax}

\bibitem[Colombani(2008)]{Colombani2008}
Colombani,~J. {Measurement of the pure dissolution rate constant of a mineral
  in water}. \emph{Geochimica et Cosmochimica Acta} \textbf{2008}, \emph{72},
  5634--5640\relax
\mciteBstWouldAddEndPuncttrue
\mciteSetBstMidEndSepPunct{\mcitedefaultmidpunct}
{\mcitedefaultendpunct}{\mcitedefaultseppunct}\relax
\EndOfBibitem
\bibitem[Jousse \latin{et~al.}(2005)Jousse, Jongen, and Agterof]{Jousse2005}
Jousse,~F.; Jongen,~T.; Agterof,~W. {A method to dynamically estimate the
  diffusion boundary layer from local velocity conditions in laminar flows}.
  \emph{International Journal of Heat and Mass Transfer} \textbf{2005},
  \emph{48}, 1563--1571\relax
\mciteBstWouldAddEndPuncttrue
\mciteSetBstMidEndSepPunct{\mcitedefaultmidpunct}
{\mcitedefaultendpunct}{\mcitedefaultseppunct}\relax
\EndOfBibitem
\end{mcitethebibliography}

\begin{tocentry}
%For Table of Contents Use Only\\
%Title: Microfluidic control of nucleation and growth of CaCO$_3$\\
%Lei Li, Jesus Rodriguez Sanchez, Felix Kohler, Anja R{\o}yne, Dag Kristian Dysthe\\
\includegraphics[width=3.5in]{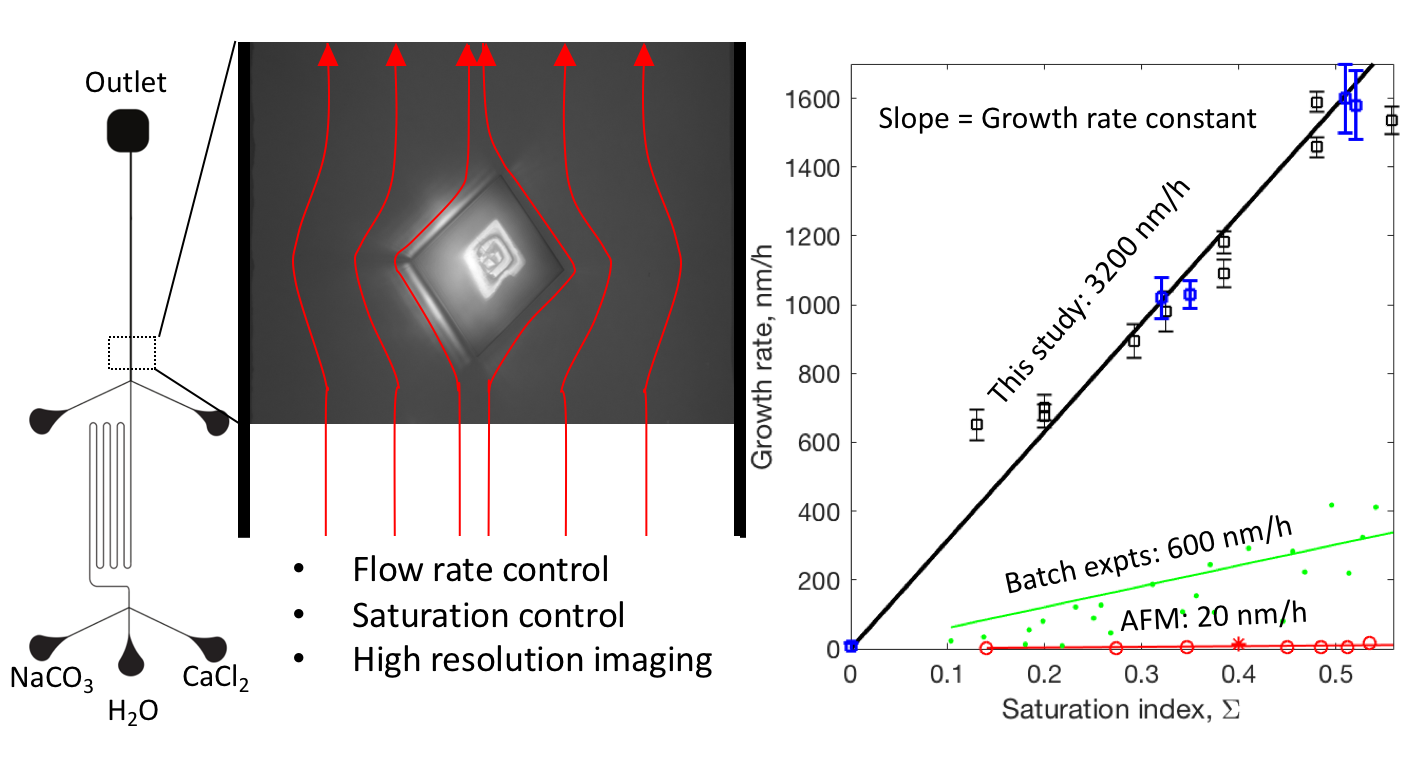}
Calcite is important in Nature and industry. We present a new method for measuring calcite crystal growth rates from solution with precise control of flow, saturation and whole crystal size. We find that previous research using AFM to calculate calcite growth rate underestimate the growth rate constant by two orders of magnitude. 
\end{tocentry}

\end{document}

\pagestyle{myheadings}
%\markboth{}{In preparation for Crystal Growth and Design}

\section{Experimetal details}
\begin{figure*}[ht]
\includegraphics[width=16cm]{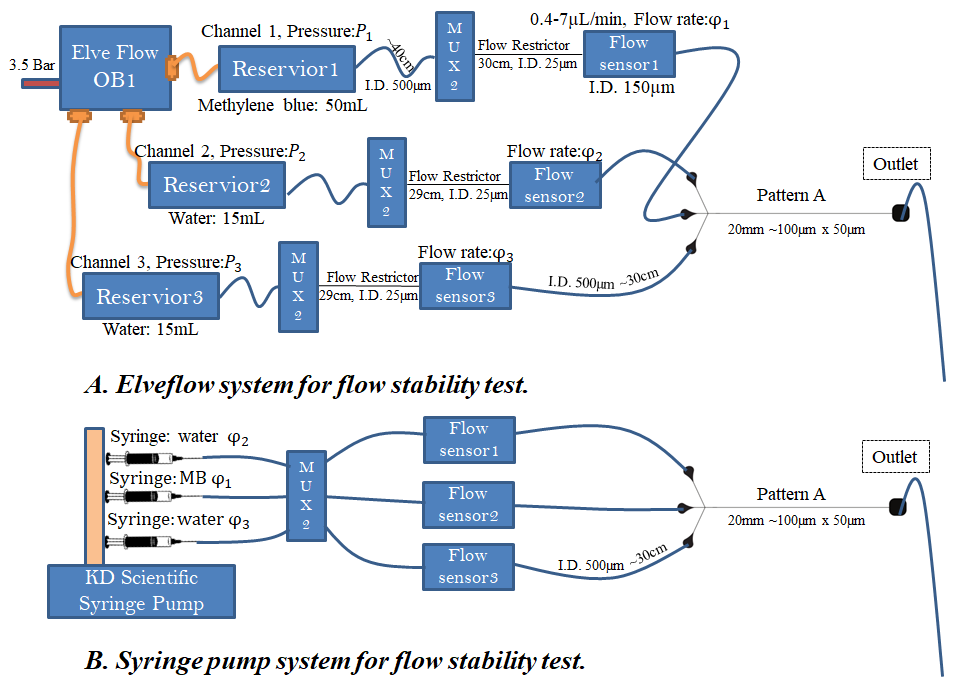}
\caption{Detailed diagram of flow rate control system and microfluidic chip for flow stability tests.}
\label{fig:flowstab}
\end{figure*}

\begin{figure*}[ht]
\includegraphics[width=16cm]{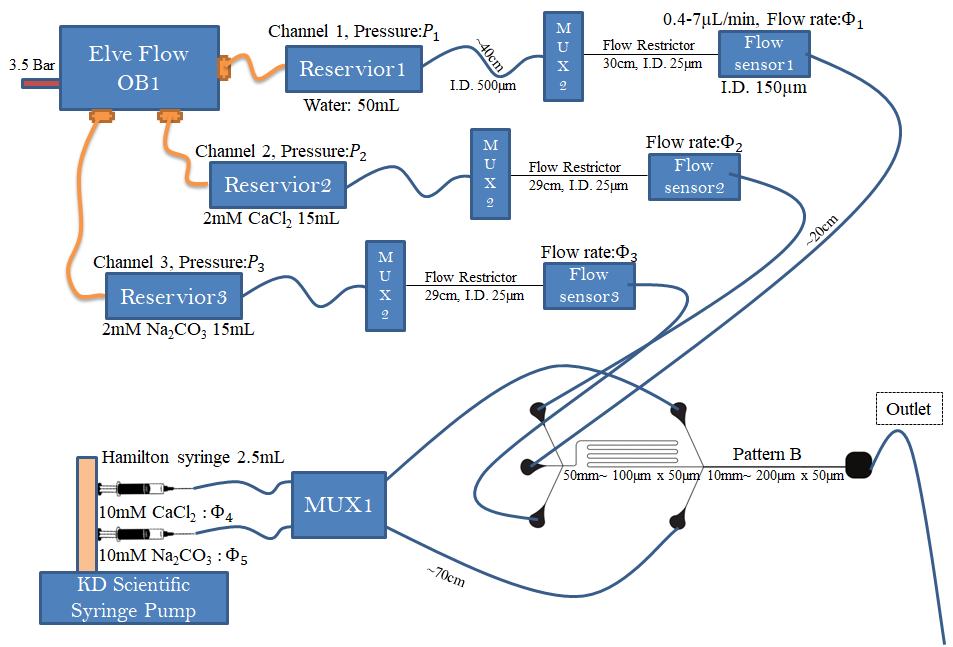}
\caption{Detailed diagram of flow rate control system and microfluidic
  chip for nucleation and growth experiments.}
\label{fig:flonucl}
\end{figure*}

Figures~\ref{fig:flowstab} and~\ref{fig:flonucl} show the detailed
setup for microfluidic flow control.

\section{Diffusion and reaction control of growth rates}
We follow Colombani~\cite{Colombani2008} that analysed the effect of
hydrodynamic boundary layers on calcite dissolution. Colombani uses
the results of Jousse et al~\cite{Jousse2005} for the hydrodynamic
boundary layer thickness, $\delta$:
\begin{equation}
\delta=2.95\left(\frac{\mu}{\rho}\right)^{1/6}D^{1/3}\sqrt{l/u},
\end{equation}
where $\mu$ is the viscosity, $\rho$ the density, $D$ the diffusion
coefficient, $u$ the far field velocity and $l$ is the length along
the crystal surface.

For pure reaction control of crystal growth the fluid concentration at
the crystal surface, $c_s$, equals the imposed concentration, $c$, far
from the surface, $c_s=c$ and the reaction controlled growth rate is
\begin{equation}
v_r=k_c(c/c_{sat}-1),
\end{equation}
where $c_{sat}$ is the saturation concentration and $k_c$ the growth
rate constant with dimension $[k_c]$=m/s ($k_c'=k_c\bar{V}$ where
$[k_c']$=mol/m$^2$/s and $\bar{V}$ is the molar volume of calcite). We now
calculate the growth rate, $v_d$, limited by diffusion through a
hydrodynamic boundary layer.  The balance of diffusion flux and growth
flux (see Equation 1 in~\cite{Colombani2008}) yields an expression for
the diffusion limited growth rate:
\begin{equation}
v_d=\frac{sD(c_s-c)\bar{V}}{s_c\delta}=k_c(c_s/c_{sat}-1),
\end{equation}
where $c_s$ is the concentration at the crystal surface, $s_c$ is the
area of the crystal, $s$ is the area of the hydrodynamic boundary
layer (see Figure 1 in~\cite{Colombani2008}) and $D$ is the diffusion
coefficient. From this one finds the concentration at the crystal
surface:
\begin{equation}
c_s=\frac{\gamma c_{sat}+c}{1+\gamma},\hspace{0.5cm}
\gamma=\frac{k_c\delta s}{Dc_{sat}s_c\bar{V}}
\end{equation}
and the ratio between the measured, diffusion limited and the purely
reaction limited growth rates is
\begin{equation}
\frac{v_d}{v_r}=\frac{c_s/c_{sat}-1}{c/c_{sat}-1}=\frac{1}{1+\gamma}.
\end{equation}
We may evaluate this for our experiments with a crystal of about
$l\sim$10~$\mu$m, $u\sim 10^{-2}$~m/s, $\mu/\rho\sim 10^{-6}$~m$^2$/s,
$D=10^{-9}$~m$^2$/s to get a boundary layer of $\delta$=9~$\mu$m. Since
$\delta\sim l$ the ratio of the  boundary layer surface to the crystal surface
is $s/s_c\sim 4$ and $k_c=2700$~nm/h~=$7.5\cdot 10^{-9}$~m/s, then
$\gamma\sim 1.5$ and $v_d/v_r\sim 0.4$.

\bibliography{calcite_growth}